\documentclass[journal]{IEEEtran}
\usepackage{amsmath,amsfonts}
\usepackage{algorithm}
\usepackage{algpseudocode}
\usepackage{array}
\usepackage[caption=false,font=normalsize,labelfont=sf,textfont=sf]{subfig}
\usepackage{textcomp}
\usepackage{stfloats}
\usepackage{url}
\usepackage{verbatim}
\usepackage{graphicx}
\usepackage{cite}
\usepackage[dvipsnames]{xcolor}
\usepackage{hyperref}
\newcommand{\cmmnt}[1]{}
\begin{document}

\title{Deep Reinforcement Learning for Trajectory Path Planning and Distributed Inference in Resource-Constrained UAV Swarms\\
}

\author{
    \IEEEauthorblockN{Marwan Dhuheir, Emna Baccour, Aiman Erbad,~\IEEEmembership{Senior Member,~IEEE},
    Sinan Sabeeh Al-Obaidi,
    Mounir Hamdi,~\IEEEmembership{Fellow,~IEEE,}}
    \thanks{This work was supported by NPRP grant \#NPRP13S-0205-200265 from the Qatar National Research Fund (a member of Qatar Foundation).  (Corresponding author: Aiman Erbad).}
    \thanks{Marwan Dhuheir, Emna Baccour, Aiman Erbad, and Mounir Hamdi are with the College of Science and Engineering, Hamad Bin Khalifa University, Doha, Qatar (e-mail: madh28916@hbku.edu.qa; ebaccourepbesaid@hbku.edu.qa; aerbad@hbku.edu.qa; mhamdi@hbku.edu.qa).}
    \thanks{Sinan Sabeeh Al-Obaidi is with Barzan Holdings QSTP LLC, Doha, Qatar. (e-mail: ssabeeh@me.com).}
}

\markboth{IEEE Internet of Things Journal}%
{Shell \MakeLowercase{\textit{et al.}}: A Sample Article Using IEEEtran.cls for IEEE Journals}

\IEEEpubid{0000--0000/00\$00.00~\copyright~2021 IEEE}

\maketitle

\begin{abstract}
The deployment flexibility and maneuverability of Unmanned Aerial Vehicles (UAVs) increased their adoption in various applications, such as wildfire tracking, border monitoring, etc. In many critical applications, UAVs capture images and other sensory data and then send the captured data to remote servers for inference and data processing tasks. However, this approach is not always practical in real-time applications due to the connection instability, limited bandwidth, and end-to-end latency. One promising solution is to divide the inference requests into multiple parts (layers or segments), with each part being executed in a different UAV based on the available resources.

Furthermore, some applications require the UAVs to traverse certain areas and capture incidents; thus, planning their paths becomes critical particularly, to reduce the latency of making the collaborative inference process. Specifically, planning the UAVs trajectory can reduce the data transmission latency by communicating with devices in the same proximity while mitigating the transmission interference.

This work aims to design a model for distributed collaborative inference requests and path planning in a UAV swarm while respecting the resource constraints due to the computational load and memory usage of the inference requests. The model is formulated as an optimization problem and aims to minimize latency. The formulated problem is NP-hard so finding the optimal solution is quite complex; thus, this paper introduces a real-time and dynamic solution for online applications using deep reinforcement learning. We conduct extensive simulations and compare our results to the-state-of-the-art studies demonstrating that our model outperforms the competing models.
\end{abstract}

\begin{IEEEkeywords}
Optimization; latency; path planning; distributed inference; CNN network; classification decision; UAVs.
\end{IEEEkeywords}

\section{Introduction}
\IEEEPARstart{R}{ecently}, 
Unmanned Aerial Vehicles (UAVs) have been widely used in various applications ranging from surveillance \cite{bejiga2017convolutional}, search and rescue operations \cite{al2018survey}, goods delivery \cite{sawadsitang2018joint} to face detection \cite{kalatzis2018edge}, delivery of wireless services \cite{kanistras2013survey}, and precision agriculture \cite{wu20205g}, which highly aroused the attention of researchers and manufacturers. UAVs outperform traditional technologies in terms of many factors such as high maneuverability, low cost, easy control system designing, and monitoring from low-level altitudes \cite{padro2019comparison}. 
Specifically, UAVs can provide architecture flexibility as their system configuration is easy, making UAVs an alternative in\cmmnt{flexible related} applications like reconnaissance missions. In addition, the ease of directing UAVs, monitoring from low-level altitudes, and capturing data from different angles give them the advantage of being used in areas that are difficult to reach by traditional methods.

Some UAV applications require more than one device to accomplish their mission with high accuracy. Furthermore, as tasks become more complicated, such as tracking an object on the ground, it becomes challenging for a single UAV to capture high-resolution images and process them due to the limited capabilities (including computational capacity and memory usage). Hence a swarm of UAVs is required in order to allow devices to collaborate in achieving a specific goal. Multi-UAV-based applications have started to appear in a plethora of applications such as oil/gas offshore inspection, delivering wireless communication in more dense areas, military-based operations, and delivering Covid-19 vaccine \cite{wu20205g}. One critical application of UAVs is surveillance and monitoring an area for many purposes, such as forest fire detection, border monitoring, and collecting data from a disaster scene.
\IEEEpubidadjcol
Moreover, a swarm of UAVs can handle more complex operations effectively. In particular, all UAVs collaborate to achieve the mission. Unlike satellites capturing images from the ground, the UAV-enabled surveillance systems can capture high-resolution pictures that are more practical for classification tasks \cite{bashmal2018learning}.

Deep neural networks (DNNs), which are employed in computer vision to analyze data collected by UAVs, have significantly improved in recent years \cite{al2018survey}. Moreover, DNNs have become a state-of-the-art solution for image classification and recognition \cite{yang2020offloading, chriki2019uav}. Although DNN has accomplished significant improvements in complicated UAV missions, one UAV cannot execute the DNN model onboard due to its resource constraints, which leads to impractical execution time \cite{huang2019deep}. In fact, these DNN networks consist of millions of neurons and billions of connections and consume large amount of resources to process the required inferences, and one UAV cannot accomplish the mission with minimum latency. Existing works tried to reduce the DNN network size, but the accuracy was affected. Hence, distributed inference in multi-UAVs is a promising solution to utilize DNN networks locally. In other words, the complete process is achieved onboard without resorting to remote servers. In order to overcome the resource-constrained problem,\cmmnt{a strategy of using deep learning has been introduced in the literature. According to this strategy,} the DNN model is partitioned into various parts/tasks, and each task is assigned to one UAV in the swarm.
Each UAV executes its dedicated task and shares the output with the next participant until obtaining the final prediction \cite{zhao2018deepthings,baccour2022pervasive}. However, working with UAVs to accomplish a mission adds a complexity of mobility and interference, represented by the data rates and the distance between devices. Previous works did not consider studying the UAVs' trajectory path and its effect on the latency of making the final prediction. Furthermore, doing in-site classification by moving devices exposed to path loss, interference, and possible disconnection has not been investigated in the literature. As a result, to allow deep neural to be processed in UAV systems, the DNN distribution should be designed to consider the hardware and physical limits, as same as intended UAVs path planning and latency of decision-making, which is one of our aims in this paper. 

Planning the UAVs' trajectory is crucial in surveillance-based applications to mitigate interference and improve the overall system performance. Different applications are studied where the UAV track is planned while executing a mission. In \cite{chang2021multi, jeong2017mobile}, authors used UAVs to provide computation offloading to User Elements (UEs) on the ground while trajectories are planned to cover specific users. The work in \cite{chang2021multi} focused on designing the UAVs' trajectory to collect data from the Internet of Things (IoT) devices. Our model involves planning the UAVs' trajectory in surveillance-based applications and collaborative image processing. In this context, we plan the route of the UAVs by dividing the surveyed area into equal-spaced cells, and we assume that each UAV covers one cell in the same time slot. Participants' mobility affects the latency of calculating the joint image classification decisions. In particular, when participants are close to each other, the interference increases, which affects data transmission. On the other hand, when UAVs are too far away, the latency to exchange data becomes very high, which makes the inference distribution not practical. To summarize, the distance between cooperating UAVs should be well planned to execute the inference request with minimum latency and low interference.

In this paper, we study the deployment of Convolutional Neural Networks (CNNs) within a surveillance UAV system. We build a collaborative strategy that distributes the computation of different CNN layers into multiple UAV devices. Particularly, the complete CNN is divided into parts/tasks, and these parts are distributed to be computed in the connected UAVs, aiming to minimize the latency while considering the resource constraints and the distance between participants. Particularly, if any participant runs out of resources, it delegates the subsequent layer to the next UAV to execute it. In addition, the trajectory of UAVs is planned to control the distance between UAVs, which highly impacts data transmission and consequently impacts the optimal placement of layers within UAVs. During this distribution, the UAVs may be requested to visit predefined hot cells. We validate our model by considering different CNN networks and different resource capacities of participants. To the best of our knowledge, we believe that our work is the first to plan UAVs' trajectory to assist a CNN distribution. Our paper's contributions are stated as follows:
\begin{itemize}
    \item We introduce a system model consisting of a swarm of UAVs capturing images and feeding them to the CNN model for image classification.
    The main objective is to minimize the time of making the final classification decision by planning their trajectory. The connected UAVs collaborate to reduce the final classification latency and finish the online requests within the UAVs.
    \item We formulate our UAVs' collaborative inference and path planning as a non-linear optimization problem that seeks to minimize the latency of making the final classification decision while planning the UAVs paths to cover the whole area by visiting different locations and controlling critical spots. This optimization respects the UAVs constraints, such as memory usage and computational complexity.
    \item We relax the optimization problem by fixing one variable and running the optimization for the other one for the maximum of possible values and choosing the value that gives the minimum latency.
    \item The formulated optimization problem is NP-hard, and it is not adequate for online resource allocation; hence we introduce a Reinforcement Learning (RL) algorithm that seeks to find the trajectory path of UAVs while distributing CNN tasks. The RL approach is solved using an efficient algorithm, namely Proximal Policy Optimization (PPO).
    \item We evaluate our system model through extensive simulations to prove its performance by considering different CNN networks, different device configurations, and different numbers of participants. Additionally, we compare our model with two benchmark system models that use different algorithms.
    \item To the best of our knowledge, our work is the first to plan the UAVs' trajectories while considering the latency of executing the distributed CNN tasks by considering the distances between cooperating devices and their effect on the interference. 
\end{itemize}

This paper is organized as follows: Section \ref{related_works} presents the related works, and section \ref{expl_System_Model} shows the system model, the problem formulation, and the reinforcement learning model. Section \ref{Perfomrance_evaluation} presents the performance evaluation and simulation results of our approach. Finally, section \ref{Conclusion} presents the conclusions.

\section{related works}
\label{related_works}
In this section, we present a literature review of distributed inference and path planning of multi-UAVs while being on a mission.
\subsection{Distributed Inferences Systems}

The distributed CNN networks have become widely known because they assist resource-constrained machines and reduce the time execution of classifying input data. Due to the exceptional performance of the distributed systems, they have been integrated into many applications, including image recognition and image classification. The work by Teerapittayanon et al. in \cite{teerapittayanon2017distributed} suggests that UAVs work as a relaying device that captures images and frames from the ground. Then the data is sent to MEC servers to process them and send the final classification decision back to the connected UAVs. This scenario depends on high-performance machines to process the data.  
Another technique was presented by Yang et al. in \cite{yang2020offloading, yang2019intelli}  in which the distribution of the model is based on the network quality, quality of the captured images, and layers. More specifically, the first segment is executed in the data-generating UAV, while the rest is offloaded to the MEC server. The partition point depends on the network quality and the quality of the images that determine the size of the DNN network.
However, this approach does not consider the latency of the transmitted data, which affects the efficiency of finding the final classification decision.

Another mainstream direction is to distribute the DNN model within the available resource-constrained devices (e.g., UAVs) in a way to accomplish the whole inference and make the final classification decision within the UAVs, without resorting to remote servers.
Disabato et al. in \cite{disabato2019distributed} proposed to distribute the CNN layers to the connected IoT devices. The classification decision was made locally while considering the connected IoT devices' memory usage and computational complexity. However, this work has not investigated the distribution of incoming online requests, which is a challenging scenario in surveillance systems. The authors in \cite{jouhari2021distributed} have investigated using multi-UAVs for surveillance and monitoring tasks in online scenarios. They focused on minimizing the latency of making the final classification decision by finding the optimal layer placement within UAV participants. However, the author considered a static path planning that does not take into account the allocation strategy and the latency of inference cooperative computation. In this paper, we propose a joint allocation and trajectory strategy that synchronizes the path of UAVs with the required data sharing in order to minimize the latency of distributed tasks.
In our previous work \cite{9498967}, we studied an online scenario of distributing the CNN layers into a swarm of UAVs, i.e., each UAV is dedicated to executing a part of the request. We also assumed in our previous work a constant data rate, i.e., the mobility of UAVs was assumed to be homogeneous, and the UAVs move while maintaining the same distance between them. Nevertheless, UAVs' heterogeneous mobility is a more realistic scenario where the distances between devices change, which affects the latency of collaborative services presented by UAVs. Therefore, in this paper, we consider a heterogeneous scenario of moving UAVs where the data rates change based on UAV distances, and the allocation of different CNN layers and path planning are chosen while mitigating the interference and minimizing the latency.

\subsection{UAVs Path Planning}

Planning the path of UAVs has been investigated by many researchers. Demiane et al. \cite{demiane2020optimized} have proposed an approach where the monitored area is divided into equal cell sizes. The covered area is classified based on the necessity of UAVs to pass through these cells, i.e., some cells do not witness many incidents, and others are critical; hence the UAVs need to traverse regularly.
In this work, the system model deals with one UAV to cover the whole area, which is not sufficient to monitor a big area and provide services. Another approach is proposed by Samir et al. in \cite{samir2019uav}, where the UAVs collect data from resource-constrained IoT devices on the ground and send them to a remote server to process them while optimizing the paths to cover the whole area. This model depends on the environment and quality of the links between the UAVs and the ground station. Hend et al. in \cite{9498662} introduced another scenario of trajectory path planning. In this approach, strategic locations are defined, and the UAVs plan their trajectories based on these strategic locations while consuming less energy. However, this study considers independent devices, and they do not present any service that requires the collaboration of UAVs in which the main goal is to plan the UAVs' path without involving any computational task. In our scenario, each UAV is supposed to capture images/collect data from the covered area, set up a CNN inference, and distribute it to the connected UAVs depending on their available resources while minimizing the classification latency and planning their trajectory paths. The UAVs are supposed to visit the crucial cells more often than other cells, which we call hot cells. The main goal of this surveillance system is to monitor the hot cells in the environment; therefore, we ensure that the hot cells are visited at each time step. Our system model obligates the UAVs to visit the hot cells each time step $t$ and satisfy part of the service demand because these hot cells are more susceptible to accidents on the spot. We highlight here that the number of UAVs should be at least equal to or greater than the number of hot cells; hence each hot cell should be monitored at each time step $t$. However, due to the different resource requests and limited resource capacities, the UAVs move to delegate the subsequent tasks to devices with available resources and minimize the delay of the distributed inferences.

The existing literature on adopting UAVs for different missions while planning the UAVs paths is investigated in many studies, such as those in \cite{chang2021multi}, and \cite{jeong2017mobile}, which use UAVs as base stations to provide services to the ground users while planning the UAVs' trajectory paths. Another approach found in \cite{9457160} uses multi-UAVs as base stations to provide services to clusters containing ground users while they navigate to maximize the coverage of the served users. Other kinds of literature plan the UAVs' trajectories in order to achieve monitoring and surveillance missions in order to make the UAVs cover the whole area \cite{yang2020offloading}. Jouhari et al. in \cite{jouhari2021distributed} investigated the distribution of CNN layers into multiple UAVs while considering already planned paths of UAVs to allocate the layers and minimize the latency of CNN execution. In our study, we are dealing with a different context compared to \cite{yang2020offloading}, \cite{chang2021multi}, and \cite{jeong2017mobile},  which is CNN distribution for monitoring mission, and we optimize the UAVs paths to assist the resource allocation and minimize the latency of the inference, which is not the case of \cite{jouhari2021distributed}.
Multi-UAVs with edge computing and path planning have been investigated in many research studies. Almost all the studies consider UAVs as an edge computing device while planning their trajectory. Huan et al. in \cite{chang2021multi} used a reinforcement learning-based model to plan the paths of UAVs until reaching their destination. Meanwhile, these UAVs provide services to terminal users on the ground, playing the role of edge devices.
Another approach was introduced by Qian et al. in \cite{liu2020path}. This approach also used reinforcement learning to draw UAVs' paths and provide edge computing services to ground users. The authors adopted the linear-based algorithm to conduct this process and evaluate the Quality of Service (QoS) for the terminal users who received UAVs' services.  

Different reinforcement learning algorithms are used to plan the trajectory of the UAV swarm. The agents generally learn the optimal policy based on the reward and penalties they receive after selecting the path planning actions. 
Another technique presented in \cite{moon2021deep,li2020path} used Deep RL (DRL) to plan the trajectory paths for tracking an object in the ground in the presence of obstacles. The authors in \cite{moon2021deep} focused on tracking the object in the presence of interference that comes from Line of Sight (LoS) and Non-LoS (NLoS); while the authors in \cite{li2020path} studied the usage of long short-term memory networks to approximate the states in the environments. All the presented studies in \cite{li2020path,moon2021deep,pan2021deep} use multi UAVs to accomplish a mission while planning the trajectory of UAVs in the presence of interference. However, they do not study the effect of the distance between the connected UAVs on trajectory planning which affects the interference and the execution time of predicting the targets.
Some studies focused on studying the trajectory of UAVs while collecting data from ground users, and they used different deep RL algorithms. The work in these studies evaluates the minimum completion time of collecting data \cite{wang2021trajectory, nguyen20223d, chen2021joint}. Other works studied the trajectory planning of UAVs while on a mission by using machine learning-based algorithms \cite{cheng2021machine, ding2021trajectory,luong2021deep,cheng2019space,zhou2020deep}. Although the previous works \cite{wang2021trajectory, nguyen20223d, chen2021joint,cheng2021machine, ding2021trajectory,luong2021deep,cheng2019space,zhou2020deep} studied the trajectory of UAVs while being involved on a mission, they did not consider studying the effect of trajectory on latency and interference, the effect of distances between UAVs on the interference while UAVs are moving, and the effect of UAVs’ capacities on the latency calculations, which are the aim factors addressed in this paper.

\section{Distributed CNN and Path Planning on Resource-Constrained UAVs}
\label{expl_System_Model}
This section introduces our system model for incorporating distributed CNN networks and planning the path of UAVs. We formulate our system model as an optimization problem that seeks to find the minimum latency between capturing an image from the ground and providing the final classification decision. The UAVs are moving to monitor a specific area while a CNN network processes the incoming requests. Because the formulated problem is NP-hard, we introduce an RL system model intended to plan the UAVs' trajectory path while capturing images from the ground to be classified.

\subsection{System Model}
\begin{figure}[!ht]
    \centering
    \includegraphics[width=0.45\textwidth]{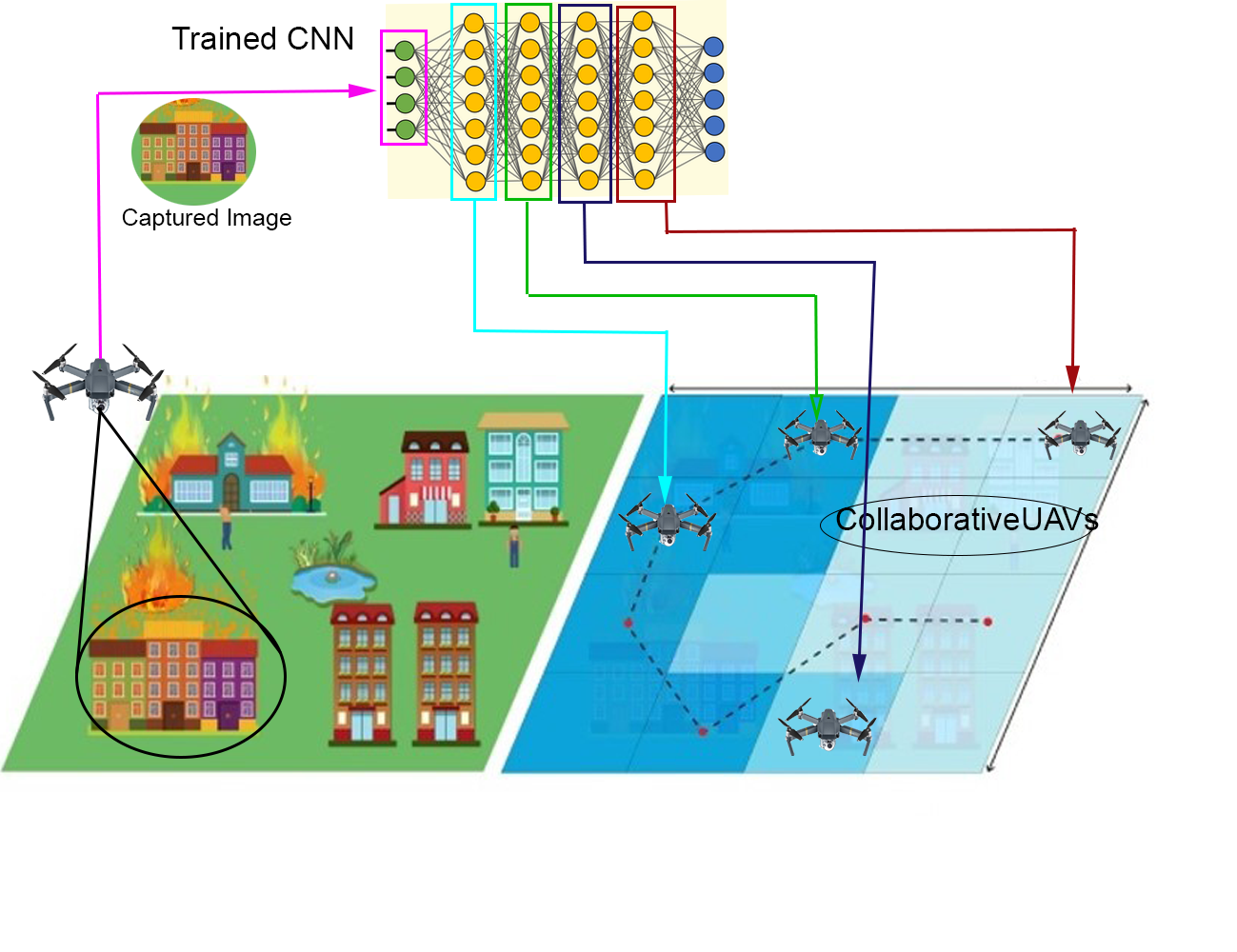}
    \caption{System Model.}
    \label{fig:graph1}
\end{figure}

Our system model is shown in Figure \ref{fig:graph1}. In this system, we divide the covered area into equal cell sizes, and the number of cells in the grid is denoted as $C$. 
Each UAV captures multiple images in real-time and collaborates with other UAVs to compute the classification prediction of the captured images.
In this scenario, the UAVs pass through the whole area; however, not all the cells have the same importance to visit, as shown in Figure \ref{fig:graph1}. Some areas are important and require UAVs to visit more than others (e.g., expected fire areas). Therefore, our system model is designed to surveil the area while focusing on the hot cell locations. We assume that the number of UAVs needs to be greater than or equal to the number of hot cells to guarantee that at least one hot cell area is visited at each time step. For non-hot cell areas, the UAVs can visit and capture requests from the visited areas. While the UAVs move in the area, they need to satisfy two essential tasks, one is executing part of the requests based on their capabilities, and the other task is to plan their path in a way that minimizes the latency of calculating the final classification prediction. At the same time, the UAV needs to mitigate interference by locating itself in a suitable position based on its distance from the rest of the UAVs in the swarm. As the energy and capacities of UAVs are limited, planning their paths is important to exploit their resources more efficiently.
If resources are not sufficient to process the entire CNN inference, the UAV sends the output of the intermediate layer to other UAVs to process the rest of the tasks. The monitored area might include different surveillance systems such as forest fire detection, tracking an object on the ground, borders monitoring, etc., which means multiple CNN models can be used according to the surveillance objectives.

The size of images captured by different UAVs is the same; however, the model can handle images of different sizes (e.g., based on the capturing camera). The processing time of the CNN depends on the type and architecture of the CNN network being used to process the data. Moreover, the distribution system in our approach is per layer. Each layer will be assigned to one UAV, and based on the UAV capabilities, part of the inference is processed. If one UAV does not have enough resources to accomplish its task of the CNN layer, this part will be delegated to another UAV to process it until the whole classification task is completed within the swarm. We highlight that for each request, all layers are computed following the trained CNN model without applying any early-exit strategy.\\
Furthermore, the latency is defined as the delay from capturing an image from the ground until making the final classification decision, and it is composed of two delays, computational and communication delays. The latency depends on the distribution of layers and the distances between UAVs. We notify that the latency is impacted by the capacity of devices assigned to different layers and the distances/data rates between the communicating devices, which depend on the path planning. The communication latency is impacted by the number of subsequent layers assigned to each UAV device. The computation latency of the same type of CNN is not the same for all requests, which depends on the number of subsequent layers and the available capacities of the devices.

The model consists of $N$ connected UAVs capturing images and feeding them to the CNN network. These UAVs collaborate to classify the input images onboard based on their capacities. The system seeks to find the suitable set of UAVs that can achieve the best latency, i.e., minimizing the time of making the final classification decision. We consider several time frames in which each frame ( e.g., a few minutes.) is divided into $T$ equal equal-time slots with a fixed length (e.g., a few seconds). Let us assume that the $i$-th UAV unit $i\in N$ is characterized by two important constraints, maximum memory usage $\bar{m_{i}}$, and maximum computational capacity $\bar{c_i}$. In this model, each UAV $ i \in \{1,...,N\}$  is assumed to generate $RQ^t_i$ requests per time step $t$ such that $ \sum_i RQ^t_i = RQ^t$, where $RQ^t$ denotes the total number of requests at $t$.

In this work, the UAVs are assumed to capture images from the ground, and the trained CNN model is used to classify the input references. Each UAV in the swarm has a copy of the trained model and accepts part of the tasks to execute it. The proposed distribution system is compatible with the typical CNN network that is designed into a pipeline of successive layers, i.e., convolutional and ReLU, and without incorporating residual blocks \cite{he2016deep}. In this context, we assume $L$ to be the number of layers of the CNN model, which represents the number of sub-tasks that are distributed to the connected UAVs to classify the input image. $j$ represents the index of the layer where $j \in\{1,..., L\}$. Each layer has a memory requirement $m_j$ and a computation demand $c_j$. The memory usage $m_{j}$ (in bytes) represents the number of weights that the layer $j$ stores multiplied by the size of the data type intended to characterize the parameters. The second requirement is the computational load $c_{j}$, which is the number of multiplications that are processed to execute the layer $j$ \cite{baccour2021rl}. Therefore, the computational load requirement for the layer $j$ is calculated as follows:
    \begin{equation}
         c_{j} = n_{j-1}\;.\;s_{j}^2\;.\;n_j \;.\;z_{j}^2
        \label{1st_eqn}
    \end{equation}
where $n_{j-1}$ is the number of the input channels at layer $j$, which represents the total feature maps that come from the layer $j-1$, $s_j^2$ and $z_j^2$ are the spatial filter size of the $j$-th layer and the spatial size of the output map, respectively. The computational load calculations of the fully-connected layer $j$ are the number of neurons $n_j$ of layer $j$ multiplied by the number of neurons $n_{j-1}$ of the layer $j-1$, and it is represented as follows:
    \begin{equation}
         c_{j} = n_{j-1}\;.\;n_j
        \label{2nd_eqn}
    \end{equation}
The memory requirements of layers are calculated as the number of weights $W_j$ of the current layer $j$ by the number of bits $b$ dedicated to storing the weight, and it is represented as:
    \begin{equation}
         m_{j}^{N_c} = W_j\;.\;b
        \label{3rd_eqn}
    \end{equation}
All UAVs are assumed to transmit a fixed power $P$ to each other in which the received power at one UAV is $P_{i,k}^t = h_{i,k}^t. P$, where $h_{i,k}^t$ is defined as the channel gain from UAV $i$ to UAV $k$ in the same time slot $t$. The channel gain is calculated as follows:
    \begin{equation}
    \small
         h_{i,k}^t = \sum_{v=1}^{C}\sum_{q=1}^{C}
          \frac {h_0}{d(v,q)^2} \; . \; \Psi_{t,i,q} \; . \; \Psi_{t,k,v}, \; \forall i \in N, \forall t \in T
        \label{4th_eqn}
    \end{equation}
where $h_0$ is defined as the median of the mean path gain at reference distance $d_0 = 1$ m, $d(q,v)$ represents the distance between the cell $q$ that contains UAV $i$ and cell $v$ that contains UAV $k$, and $\Psi_{t,i,q}$ represents a binary variable that takes 1 if the UAV $i$ exists in cell $q$ at time frame $t$ and 0 otherwise.
When UAV $i$ sends intermediate data to UAV $k$ in time slot $t$, the achievable data rate can be calculated as follows:
    \begin{equation}
         \rho_{i,k}(t) = B_{i,k}^t\: log_{2}(1+\frac{P_{i,k}^t}{\sigma^2})
        \label{6th_eqn}
    \end{equation}
where $\sigma^2$ is the thermal noise power, and $B_{i,k}^t$ is the transmission bandwidth between UAV $i$ to UAV $k$ at time slot $t$. We note that the Doppler effect is out of the scope of this paper, and it will be considered in future work.

\subsection{Problem Formulation}
Our system model is designed to deal with real-time requests, find the optimal placement of the layers in the available UAVs, find the optimal position of the connected UAVs, and plan the path they can follow. The UAVs should avoid colliding with others and provide monitoring services to the covered area.
We formulate this methodology as an optimization problem that seeks to find the minimum latency of calculating the classification prediction and plan the UAVs' trajectory to accomplish their mission of monitoring a region.

The proposed optimization problem depends on two decision variables $\Psi_{t,i,q}$ and $\delta_{t,r,i,j}$, and they are defined as follows:

$\forall r\in RQ^t,\forall i,k \in N, \forall j\in L$

    \begin{equation}
    \begin{aligned}
    \small
        \delta_{t,r,i,j} 
        = \left\{ 
        \begin{array}{l}
            1 \;\text{if UAV $i$ executes the layer $j$ of request $r$}\\
            \text{at time frame $t$}\\
            0 \; \text{otherwise}\\
        \end{array}
        \right.
        \end{aligned}
        \label{7thh_eqn}
    \end{equation}
    \begin{equation}
        \small
        \Psi_{t,i,q} 
        = \left\{ 
        \begin{array}{l}
            1 \;\text{if the uav $i$ exists at cell $q$ at time instance $t$} \\
            0 \; \text{otherwise}\\
        \end{array}
        \right.
        \label{5th_eqn}
    \end{equation} 
The objective function minimizes the latency of executing all decision variables at time frame $T$, which is shown in equation (\ref{8th_eqn}). We emphasize that we adopted the method of per-layer distribution as it is suitable for systems with a limited number of UAVs and compatible with the UAV environment that we described in the system model. In addition, the per-layer distribution leads to less data transmission, hence decreasing the energy consumption as same as reducing the dependency between UAV devices. Finally, to support the dynamic of the model (e.g., change of requests rate, change of the number of UAVs joining or quitting for charging.), the optimization process is executed periodically.

Our problem is an Integer Linear programming (ILP) optimization and it is formulated as follows:
\begin{equation}
    \begin{aligned}
    \small
        \min_{\Psi_{t,i,q},
        \delta_{t,r,i,j}}
        \sum_{t=1}^{T}
        \sum_{r=1}^{RQ^t}
        \sum_{i=1}^{N}
        \sum_{k=1, k\neq i}^{N}
        \sum_{j=1}^{L-1}
        \delta_{t,r,i,j}\;.\;
        \delta_{t,r,k,j+1}\\ \;.\;\frac{K_{r,j}}
        {\rho_{i,k}(t)} 
        \;+
        \sum_{i=1}^{N} t_i ^{(p)} \;+
        t_s
        \end{aligned}
        \label{8th_eqn}
    \end{equation}
    s.t.
    \begin{equation}\tag{8a}
    \small
        \sum_{r=1}^{RQ^t}
        \sum_{j=1}^{L} 
        \delta_{t,r,i,j} \;. \;m_{j} 
        \le 
        \bar{m_i}, \quad
        \forall i 
        \in{N},
        \label{8ath_eqn}
    \end{equation}

    \begin{equation}\tag{8b}
    \small
        \sum_{r=1}^{RQ^t}
        \sum_{j=1}^{L} 
        \delta_{t,r,i,j} \;. \; c_{j} 
        \le 
        \bar{c_i}, 
        \forall i 
        \in{N},
        \label{8bth_eqn}
    \end{equation}
    \begin{equation}\tag{8c}
    \small
        \forall r 
        \in {RQ^t}, 
        \forall j 
        \in {L},
        \forall t 
        \in{T},\;
        \sum_{i=1}^{N} 
        \delta_{t,r,i,j}  
        = \left\{ 
        \begin{array}{l}
            1 \;\text{\;if $j$ $ \le $ $L$}\\
            0 \; \text{\;otherwise}\\
        \end{array}
        \right. \quad
        \label{8cth_eqn}
    \end{equation} 
      \begin{equation} \tag{8d}
        \forall t 
        \in {T}, 
        \forall j 
        \in {L},
        \forall i 
        \in {N},
        \forall r 
        \in{RQ^t} \; ,\;
        \delta_{t,r,i,j} \in \{0,1\}
        \label{8dth_eqn}
    \end{equation} 
    \begin{equation}\tag{8e}
        \forall t 
        \in {T}, 
        \forall i 
        \in {N},
        \forall v 
        \in{C} \; ,\;
        \Psi_{t,i,q} \in \{0,1\}
        \label{8eth_eqn}
    \end{equation} 
    \begin{equation}\tag{8f}
    \small
        \sum_{v=1}^{C}
        \Psi_{t,i,v} 
        = 1
        , \quad
        \forall i 
        \in{N},
        \forall t 
        \in{T}
        \label{8fth_eqn}
    \end{equation}
    \begin{equation}\tag{8g}
    \small
        \sum_{i=1}^{N}
        \Psi_{t,i,v} 
        \le 1
        , \quad
        \forall v 
        \in{C},
        \forall t 
        \in{T}
        \label{8gth_eqn}
    \end{equation}
    \begin{equation} \tag{8h}
    \small
        \sum_{i=1}^{N}
        \Psi_{t,i,q} == 1
        , \quad
        \forall q 
        \in{\Omega},
        \forall t 
        \in{T}
        \label{8hth_eqn}
    \end{equation}
and where
\begin{equation}
    \small
        t_s =
        \sum_{t=1}^{T}
        \sum_{r=1}^{RQ^t}
        \sum_{k=1,  k\neq s}^{N}
       \bar\delta_{t,r,s,1}.\delta_{t,r,k,1}. \frac{K_s}{\rho_{i,k}(t)}.
         \label{9th_eqn}
    \end{equation}
    \begin{equation}
    \small
        t_i^{(p)} =
        \sum_{t=1}^{T}
        \sum_{r=1}^{RQ^t}
        \sum_{j=1}^{L}
        \delta_{t,r,i,j} \;.
         \frac{c_{j}}{e_i}
         \label{10th_eqn}
    \end{equation}
Equation (\ref{8th_eqn}) calculates the total latency of making the final classification of the captured images, in which it consists of three main parts, and they are:
\begin{itemize}

\item 
The latency $t_s$ in equation (\ref{9th_eqn}) is the time required by the UAV that captures the request from the ground to transmit its collected data to the UAV computing the first layer of this request. $K_s$ is the size of the data collected by the source UAV $s$ from the ground. In equation (\ref{9th_eqn}), $\bar{\delta}_{t,r,i,1}$ represents the complement of the decision variable $\delta_{t,r,i,1}$.

\item $t_i^{(p)}$ is the total time required to compute all tasks assigned to device $i$, which is presented in equation (\ref{10th_eqn}). This time is defined as the ratio between the computational requirement $c_{j}$ of the layer to the number of multiplications $e_i$ that UAV $i$ can process in a second.

\item The time required to transmit the intermediate representation of the input image from the UAV device $i$ to the UAV device $k$ that are assumed to be connected in the swarm and can be represented in equation (\ref{11th_eqn}) as shown below:
    \begin{equation}
        \frac{K_{r,j}}{\rho_{i,k}(t)}
        \label{11th_eqn}
    \end{equation}
where $K_{r,j}$ is the intermediate data generated from layer $j$ of request $r$, $\rho_{i,k}$ represents the transmission data rate between the UAV device $i$ and the UAV device $k$. It represents the distance and the quality of the transmission connection between UAVs. Since UAVs are moving, the value of $\rho_{i,k}$ is changing according to the distance between the UAVs over the time interval in the optimization. The data rate is calculated according to equation (\ref{6th_eqn}) as introduced in \cite{challita2019interference}.
\end{itemize}

The constraints in equations (\ref{8ath_eqn}) and (\ref{8bth_eqn}) are added to set a threshold for the maximum memory usage and the maximum computational capacity of the UAVs in the swarm, respectively. These thresholds are set to make the UAVs work within the allowable limit that respects their resources and to make the latency calculation compatible with the real models. Moreover, the constraint in equation (\ref{8cth_eqn}) is added to ensure that each layer is computed by one UAV, and the constraints in equations \ref{8dth_eqn} and \ref{8eth_eqn} refer to the binary indication of the decision variables and they accept only the value of 0 or 1. 
The constraint in equation (\ref{8fth_eqn}) is set to ensure that each cell is visited only by one UAV at each time step.
This condition is added to avoid collision between different UAVs that cannot co-exist in the same area at the same time.
Meanwhile, the constraint in equation (\ref{8gth_eqn}) is set to guarantee that each UAV visits only one cell at each time instance $t$. The constraint in equation (\ref{8hth_eqn}) refers to the hot cells that UAVs need to traverse and capture images from them. We highlight that $\Omega$ denotes the set of hot cells in the grid. If the whole area should be monitored during the period $T$, the hot cells should include the indexes of all cells over time. Moreover, the size of hot cells must be equal to or less than the number of UAVs.

The optimization problem in equation (\ref{8th_eqn}) is NP-hard, and finding the optimal solution is quite difficult because it is time-consuming and cannot be used to receive online requests and execute real-time classifications. In addition, our optimization problem changes periodically, and it assumes full knowledge of the set of incoming inferences. The UAVs' resources should also be reinitialized after computing the classifications. Hence, using the optimization (\ref{9th_eqn}) is not practical in the case of dynamic environments. Recently, reinforcement learning has shown exemplary performance in dealing with dynamic and realistic models \cite{liu2020adaptive}. The RL systems can learn the optimal solution from the knowledge of the environment states and make the best prediction from the available choices based on the previous rewards. Moreover, RL systems have the advantage of continuous learning to adapt to any fluctuations in the dynamic system models. In summary, to distribute the online classifications among UAVs and find their optimal path, we introduce an RL-based model that can approximate the optimal solution. Table \ref{tab:data_summary} presents a summary of symbols presented in our work for ease of reference.

\begin{table*}[!t]
    \caption{List of important Key Notations}
    \label{tab:data_summary} 
    \centering
    \begin{tabular}{|p{1.5cm}|p{6.5cm}||p{1.5cm}|p{6.5cm}|}\hline
        Notation & Description & Notation & Description \\\hline
       $C$ & The number of cells in the grid & $T$ & The number of time slots \\\hline
       $L$ & The number of layers & $N$ & The UAV number  \\ \hline
       $RQ$ & The Number of requests & $n_{j}$ & The number of the input channels at layer $j$\\ \hline
       $s_j$ & The spatial filter size coming from the feature map of the $j$-th layer & $z_j$ & The spatial size of the output map\\ \hline
       $W_j$ & The weight number of layer $j$ & $b$ & The number of bits dedicated to store the weight\\ \hline
       $d(q,v)$ & The distance between the cell $q$ that contains UAV $i$ and cell $v$ that contains UAV $k$ & $h_{i,k}^t$ & The channel gain between UAVs $i$ and $k$ at time slot $t$  \\ \hline
       $P_{i,k}^t$ & The transmitted power between UAVs $i$ and $k$ at time slot $t$ & $B_{i,k}^t$ & The transmission bandwidth between UAV $i$ to UAV $k$ at time slot $t$\\ \hline
       $\rho_{i,k}^t$ & The data rate between UAVs $i$ and $k$ at time slot $t$ & $K_{r,j}$ & The size of the intermediate transmission of the output data from the layer $j$ of the request $r$ \\ \hline
       $\Psi_{t,i,v}$ & The decision variable for trajectory path planning & $t_s$ & The transmission time required by the UAV that captures images from the ground to the UAV that executes the first layer \\ \hline
       $\delta_{t,r,i,j}$ & The decision variable for allocation & $t_i^{(p)}$ & The transmission time required to compute the layers assigned to device $i$\\ \hline
       $e_i$ & The number of multiplications that UAV $i$ can process in a second & $\Omega$ & The set of hot cells\\ \hline
       $S_f$ & QoS parameter & $\phi_\Omega$ & demand service vector \\ \hline
       $\bar{c_i}$ & maximum capacity of the device $i$  & $\bar{m_i}$ &  maximum memory of the device $i$\\ \hline
       $c_j$ & The computational load requirement of layer $j$ & $m_j$ & The memory requirement of layer j\\\hline
       $S$ & Set of states & $A$ & Set of actions \\ \hline
       $R$ & Immediate reward & $P$ & The possible transition probabilities of the states \\ \hline
       $\gamma$ & Discount factor & $\alpha$ & Learning rate  \\ \hline
       $M$ & Batch size & $\pi$ & RL policy \\ \hline
       $v^{\pi}(s)$ & The expected return of the reward function after applying action $a$ on state $s$ & $\theta$ & Neural network weights \\ \hline
       $\epsilon$ & Exploration rate & $\varepsilon$ & PPO clipping range \\ \hline
       $\lambda$ & PPO coefficient & $\sigma^2$ & Thermal noise power \\ \hline
       $l_j$ &  The layer $j$ to be allocated & $N_i$ &  The UAV $i$ to be deployed \\ \hline
       $P^*_{s_i}$ &  The position of UAV $i$ in the grid & $D$ &  The number of parameters that will be included in the set states of RL model\\ \hline
       $\eta$ &  The policy probability ratio  & $R_{\Omega}$ & the immediate reward of passing through hot cells\\ \hline
    \end{tabular}
\end{table*}
\subsection{Reinforcement Learning for Trajectory Path Planning and Inferences Distribution on Resource-Constrained Swarm of UAVs}
Finding the optimal solution to equation (\ref{8th_eqn}) is complex and time-consuming due to the nonlinearity in equation (\ref{4th_eqn}) and the objective function; hence, it cannot be applied in real-time scenarios. To solve the problem, we adopted a real-time and dynamic solution, which is deep RL. The learning process of RL models is done offline, and after the convergence of the algorithm and the allocation policy is learned, the actions will be taken in real-time. Specifically, the deep DNN model that we adopt for RL has 2 layers and 64 neurons for each layer, which is very light, and the system's complexity is negligible; therefore, the RL solution can take on-the-fly actions, and our algorithm can handle the layers' allocations and plan the UAVs trajectory paths online.
Our RL system model seeks to find the set of UAV devices that collaborate to achieve the minimum classification latency and plan their paths that further contribute to minimizing this latency. This section presents our method of designing the system environment, sets of states and actions, developing a practical reward function, presenting the RL agent, and selecting the most appropriate RL technique. 

The reinforcement learning agent chooses the actions based on the environment states to maximize the long-term reward. Accordingly, the primary objective is to find the optimal policy that accomplishes the maximum reward. In the exploration phase, the agent learns by trying different action choices and receiving positive or negative feedback in terms of reward. The agent learns from trial and error by interacting with the changes in the environment and accordingly adapting to it. During this phase, the RL agent gains experience from the rewards and penalties of each action, and after the process of exploration, it becomes ready to converge to the maximum reward policy. In our platform, the RL agent is doing two tasks (allocation and path planning decisions).

At each time step, the RL agent chooses the location of each UAV and decides if this UAV can handle the current layer of the CNN. The decision taken depends on many factors that have an impact on the distribution process, such as the trained CNN network, the remaining resources of the UAVs, the location of the other UAVs in the grid, the current layer, and the positions of hot cells in the grid. Our RL-system model is represented by a  framework of Markov Decision Process (MDP) depicted by $(S, A, P, R, \gamma)$. $S$ represents the set of states, $A$ is the set of actions, $P$ is the possible transition probabilities of the states, $R$ is the immediate rewards received for each action, and $\gamma$ represents the discount learning factor. The MDP elements are explained in detail in the following sections.

\begin{figure*}[!t]
\centering
    \includegraphics[width=1\textwidth]{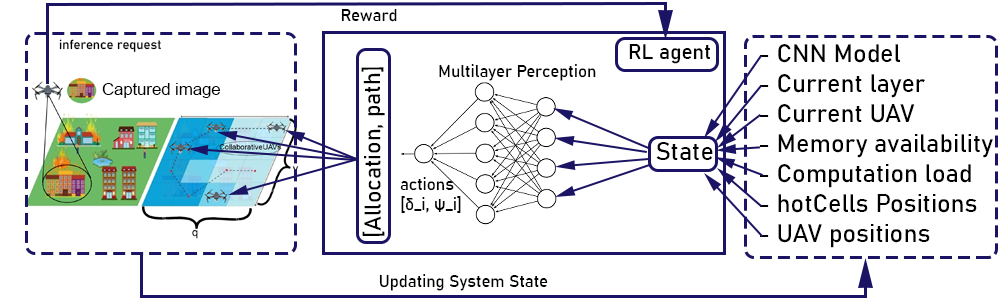}
    \caption{RL System Model.}
    \label{fig:graph2}
\end{figure*}

\begin{enumerate}
  \setcounter{enumi}{0}
  \item Environment Modeling
\end{enumerate}
The environment model of this system includes a set of UAVs collaborating to accomplish online inference requests. Let $\pi$ denote the stochastic Policy in which $\pi : S \times A \rightarrow [0, 1]$. The MDP environment model is set to receive the instructions from a centralized agent that takes actions $A_t$, receives rewards/penalties $R_t$, and learns the optimal policy $\pi^*$ from the feedback it receives at each time step $t$. The agent seeks to find the optimal policy $\pi^*$, which has the maximum value $v^{\pi^*} (s)$ for all elements $s \in S$. The value set of $v^{\pi}(s)$ represents the expected return of the reward function after applying action $a$ on state $s$ while trying to reach the optimal policy, and it is expressed as follows:
\begin{equation}
    v^{\pi}(s) = \mathbf{E}_{a_{t},s_{t+1}}\Bigg(\sum_{k=1}^{\infty} \gamma^{k-t} R_k|S_t = s\Bigg),
    \label{12th_eqn}
\end{equation}
By receiving rewards and penalties, the agent can learn from the past experiences of the multi-decisions, i.e., layers allocation and the trajectory path, in order to increase the rewards $R_t$. In our system model, one episode consists of multiple time steps. It represents a distribution of one request to the available UAVs while moving in the grid and avoiding colliding with each other. One time step $t$ represents the allocation of one layer of the CNN request in one of the UAVs and the location of this UAV. In addition, each episode is independent of other episodes, and the reward is set to $0$ at the beginning of each one. We note that our agent does not have visibility of the environment design nor information about the next state. Alternately, the agent learns the optimal policy $\pi^*$ from the interaction with the environment, receiving rewards/penalties, i.e., the optimal policy $\pi^*$ is a function to be learned by taking action $A_t$ to the current state $S_t$ in order to move the agent to the next state $S_{t+1}$, i.e., $\pi^*: S \rightarrow  A$. Figure \ref{fig:graph3} depicts the representations of the states and actions of one episode. The episode length depends on the size of the CNN and the number of UAVs, in which the length of one episode is $L \times N$ where $L$ is the number of layers and $N$ is the number of UAVs. Hence, the episode length in each CNN network is different.
\begin{enumerate}
  \setcounter{enumi}{1}
  \item States and Actions
\end{enumerate}
In order to achieve high system performance, the states should include helpful information that describes the environment sufficiently to the agent. Our system model is depicted in Figure \ref{fig:graph2}. The action vector is composed of two parts, allocation and trajectory planning, i.e., $A = [a_1, a_2]$. The set of states $S$ includes all the configurations that impact the transitions of UAVs in the grid and the computations of different CNN layers. Hence, at each time step, the information received by the agent includes the following observations:
\begin{itemize}
\item  $l_j$ : the current layer to be allocated,
\item  $N_i$ : the current UAV,
\item  $\bar{m_i}$ : the memory capacity of the UAV device $i$,
\item  $\bar{c_i}$ : the computational load of the UAV device $i$,
\item  \{$P_{s_1}^*, . . ., P_{s_N}^*$\} : the position of UAVs in the grid,
\item  \{$\Omega_1, . . ., \Omega_N$\} : the position of hot cells in the grid. 
\begin{figure}[!t]
    \centering
    \includegraphics[width=0.6\linewidth]{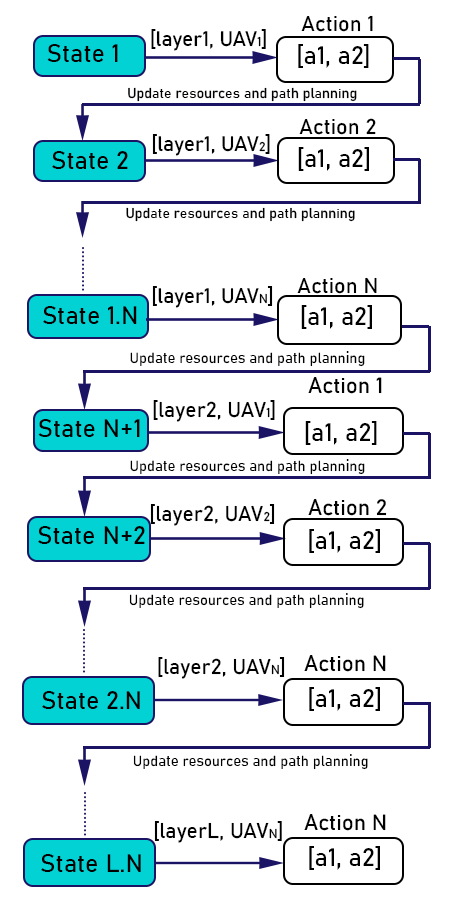}
    \caption{Episode representation of the RL Model.}
    \label{fig:graph3}
\end{figure}
\end{itemize}
The model state is hence a vector that contains the following: $\{ l_j, N_i, m_i, c_i, \{P_{s_1}^*, . . ., P_{s_N}^*\}, \{\Omega_1, . . ., \Omega_N\}\}$. We emphasize that $i$ is the index of UAV devices, and $\Omega_N$ is index of hot cells in the grid where the indices of UAVs positions and hot cells are known to the model, i.e., the length of the UAVs position vector is the length of the grid containing the positions of the UAVs in the grid. The vector is composed of binaries where 1 means that it is an index of UAV or hot cell. According to the described system design, the states set length is $(2 * size(grid) + D)$. $D$ represents the number of parameters (e.g., memory, computational load, current layer, etc.) that will be included in the states, which is equal to $4$. In addition, at the end of each time step, the memory, computational capacity, and the UAVs’ positions are updated according to the taken action. Moreover, the current layer and current UAV device are also updated according to the number of the running time step. The updated data will be the input to the next time step. In addition, the shape of the action vector consists of two parts, one is the allocation part, which is a binary index that indicates whether to allocate or not. The other part is the index of the cell where the UAV should move.

\begin{enumerate}
  \setcounter{enumi}{2}
  \item Reward function
\end{enumerate}
The reward function is an important parameter that needs to be well-designed in order to get high system performance and lead to minimum latency in computing the final prediction. In RL systems, when the agent takes action $A_t$ for the current state $S_t$, it gets feedback about the result which is represented as a reward. 
Our reward function is based on the ability of the RL agent to respect the constraints formulated in the problem (\ref{8th_eqn}).

\begin{equation}
    \begin{aligned}
    \left\{ 
    \begin{array}{l}
    \text{Cons1:\; if }  \sum_{j=N\times\lfloor(t-1)/N\rfloor+1}^t a^j_1 = 1 \qquad \text{constraint \ref{8cth_eqn}}\\
    \text{Cons2:\; if }  j \in \{N\times\lfloor(t-1)/N\rfloor+1,...,t\} \rightarrow\\ 
     a^j_2 \notin \text{UAVsVisitedList} \qquad \qquad \qquad \text{constraints \ref{8fth_eqn} and \ref{8gth_eqn}}\\ 
    \text{Cons3:\; if } a_{1}^t = 1 \rightarrow m_{j} \le \bar{m_i} \text{ and }  c_{j} \le \bar{c_i} \\  \qquad \qquad \qquad \qquad \qquad \qquad \qquad \; \; \; \text{constraints \ref{8ath_eqn} and \ref{8bth_eqn}} 
    \end{array}
        \right.
    \end{aligned}
    \label{13th_eqn}
\end{equation}

$Cons1$ is added to make the sum of actions equal to 1, and it is equivalent to the constraint in equation (\ref{8cth_eqn}). This constraint ensures that the layer related to the steps in $\{N\times\lfloor(t-1)/N\rfloor+1,...,t\}$ is assigned to only one UAV. $Cons2$ indicates that each UAV can visit one cell and can exist in one cell at each time-step, and $UAVsVisitedList$ is an empty list defined at the beginning of each layer that stores the cells UAVs visit. $Cons3$, on the other hand, refers to the maximum threshold of the UAV resources to make sure that the chosen UAV has enough resources to execute the specified layer of the request. 
Then we define the instant reward, which refers to the total reward of respecting the constraints of the multi-task altogether, i.e.,
\begin{equation}
    R_t = Cons1 * Cons2 * Cons3
    \label{14th_eqn}
\end{equation}
According to equation (\ref{14th_eqn}), if $Cons1$ is not respected, i.e., ($Cons1 = 0$), this means that the related task is rejected and cannot be handled by any of the connected UAVs due to the non-availability of the UAVs or this layer is assigned to multi-UAVs at the same time. Therefore, the reward is set to 0 as a penalty to encourage the RL system to avoid these invalid cases. Moreover, if more than one UAV exists in the same cell in the grid, i.e., ($Cons2 = 0$), the step reward will be 0, and the system recognizes this as a penalty. Finally, if $Cons3$ is not respected, i.e., ($Cons3 = 0$), this indicates that the chosen UAV does not have enough resources to handle the request. In this case, the immediate reward is equal to 0. The step reward will be equal to 1, if all the constraints are respected. In addition, the first time step, i.e., $t = 1$ is dedicated to calculating the time delay of the transmission of the image to the UAV that will compute the first layer, hence the delay time $\frac{K_s}{\rho_{s,i}(t)}$ of calculating the first layer is added as a penalty to the total reward as indicated in equation (\ref{9th_eqn}). Then, in the remaining time steps, the time delay of calculating the computation and communication latency equal to ($\frac{K_j}{\rho_{i,k}(t)} + \frac{c_j}{e_i}$) as indicated in equation (\ref{9th_eqn}) are added as a penalty to the instant reward.
  
In equation (\ref{15th_eqn}), $S_f$ is denoted to the QoS factor. Particularly, this factor refers to the encouragement level of the agent to traverse through hot cells and accept requests. According to equation \ref{15th_eqn}, when $S_f$ becomes higher, the reward $R_\Omega$ becomes high, and it encourages the RL system to visit hot cells and provide services. In other words, the UAVs should traverse through these cells and capture requests. According to this scenario, at each time step, the UAV traverses through hot cells, and each hot cell has a demand service to be fulfilled. The demand vector of all hot cells is denoted by $\phi_\Omega$.

Hence, we define the reward-related to the services that are offered to the hot cells as the same work that has been done in \cite{liu2020path}:
\begin{equation}
    R_{\Omega} = \frac{S_f}{1+\sum_{i=1}^{\Omega} \phi_\Omega (t)}
\label{15th_eqn}
\end{equation}
\begin{enumerate}
  \setcounter{enumi}{3}
  \item RL Agent Design
\end{enumerate}
The primary goal of the RL agent is to maximize rewards by experimenting with various environment states and select actions. In this way, the agent learns the model's optimal policy $\pi^*$ from its surroundings. Finding the optimal policy starts by creating a strategy of the optimal set of action-values $Q(S_t, A_t)$, which gives the agent a hint about the expected future reward of taking a specific action $A_t$. The function of action-value $Q(S_t, A_t)$ of RL-agent, which represents the future expected reward and can depict the performance/goodness of the selected action and is shown as follows:
\begin{equation}
\small
\begin{aligned}
    Q(s_t,a_t) = \mathbf{E}_{a_{t},s_{t+1}}\Bigg(\sum_{k=1}^{N} \gamma^k R_{t+k}|S_t = s, A_t = a\Bigg)
\label{16th_eqn}   
\end{aligned}
\end{equation}
where $\gamma$ is the discount factor in which $\gamma \in [0,1]$ and it represents the average of immediate reward to the long-term rewards. The differentiated equation of transitions in states and actions can be expressed as\cite{moon2021deep}:
\begin{equation}
\small
    Q(s,r) \leftarrow Q(s,r)+\alpha (R_t +\gamma\; \max_{a} \hat{Q}(s,a)
    - Q(s,a))
    \label{17th_eqn}  
\end{equation}
where $\alpha$ represents the learning factor.

\begin{enumerate}
  \setcounter{enumi}{4}
  \item PPO Algorithm
\end{enumerate}
PPO was introduced in \cite{schulman2017proximal}, which belongs to the family of policy gradient methods.
In order to improve the RL system performance and handle multi-task operations, we chose the PPO as it can relax the complex constraints and replace them with flexible ones\cite{azar2021drone}.
The PPO algorithm's main goal can be summarized as follows:
\begin{equation}
\small
\begin{aligned}
J^{PPO} = E_{s,a}[min(\eta * A^{\pi_{old}},clip(\eta,1-\varepsilon,1+\varepsilon)\\
A^{\pi_{old}})]
\end{aligned}
\label{18th_eqn}  
\end{equation}
where $\eta$ represents the policy probability ratio and is calculated as follows:
\begin{equation}
    \eta = \frac{\pi(a,s)}{\pi_{old}(a,s)}
    \label{19th_eqn}  
\end{equation}
$A^{\pi} = Q(s_t,a_t) - v^{\pi}(s)$, which refers to the advantage function that measures the quality of action compared to the actions available in the state vector and can be calculated by using different methods \cite{schulman2015high}. The value function $v^{\pi}(s)$ measures the long-term reward of being in the state $s$ following the policy $\pi$, while $Q(s_t,a_t)$ measures the long-term reward of taking specific action in the state vector $s$ following the policy $\pi$. $\varepsilon$ is denoted to the PPO clip range \cite{bohn2019deep}. In the case of $A^{\pi_{old}} > 1$, the output of the function $\eta$ becomes very large, i.e., (larger than 1) in order to make the system performance higher; however, this leads to unsuitability in the learning steps. PPO solves this problem in equation (\ref{18th_eqn}) by restricting the value of $\eta$ to be in the range of $[1 - \varepsilon, 1 + \varepsilon]$.

In our system model, the distribution of participants varies and depends on the available capacities. Moreover, the system is dynamic due to the number of UAVs joining or leaving the swarm, and the action space vector is related to the active UAV devices. Therefore, it is difficult to apply traditional reinforcement learning techniques to solve the layers allocation and planning of the trajectory path of UAVs, where Q-tables are adopted to keep the historical Q-values. We adopt Deep Reinforcement Learning (DRL) to solve the aforementioned challenges and learn the parametric representation of the action-value function $Q(s, a)$. The goal of the DRL is to minimize the loss function obtained as follows:
\begin{equation}
    L(\theta) = \mathbf{E}{[(R_t+\gamma\max_{\bar{a}}Q(\bar{s}, \bar{a}, \bar{\theta})-Q(s,a,\theta))^2]}
    \label{20th_eqn}
\end{equation}
Where $Q(s,r)$ is approximated to $Q(s,r,\theta)$ and $\theta$ represents the weights of deep learning networks.


\begin{algorithm}
\caption{RL algorithm}\label{alg:cap}
\begin{algorithmic}[1]
\small
\State \textbf{Initialization}
\State \text{Randomly initialize the parameters $\theta$ of the Q-network.}
\State \text{Randomly initialize the parameters $\theta$ old of the Q-network:} \text{$\theta_{old} \leftarrow \theta$.}
\State \text{Randomly distribute UAVs in the grid: }
\For{$each \; UAV \; in\; C$}
    \State \text{distribute all UAVs in the grid}
\EndFor
\State \textbf{PPO2 training: }
\For{each episode $r$ $\in$ $R$}
\For{each $t$ $\in$ $T$}
\For{each $j \in \{1,2,3,...,L\}$}
\For{each $i \in \{1,2,3,...,N\}$}
    \State \text{$S(i) = \{j,i,m_j,c_j,\{\Omega_1,\Omega_2,...,\Omega_N\},$}
     \text{\qquad \qquad \qquad \qquad \quad $\{P_{s_1}^*,P{s_2}^*,...,P_{s_N}^*\}\}$}
    \State \text{choose action $A_i$ based on $\epsilon$}
    \If{$Cons1$} \qquad\text{equation (\ref{13th_eqn})}
    \If{$j = 1$}
        \State \text{$R_t = R_t - \frac{K_1}{\rho_{i,k}(t)}- \frac{c_{r,1}}{e_i}$}
    \ElsIf{$j \neq 1$}
    \If{$Cons3$}\qquad\text{equation (\ref{13th_eqn})}
    \State \text{$R_i = R_i - (\frac{K_{r,s}}{\rho_{i,k}(t)} + \frac{c_{r,j}}{e_i})$}
    \State \text{$a_2 = i$}
    \EndIf
    \If{$Cons2$} \qquad\text{equation (\ref{13th_eqn})}
    \State \text{$R_t = R_t + 1 \qquad $equation(\ref{14th_eqn})}
    \EndIf
    \If{$UAV(i) \notin \Omega$}
    \State \text{$R_t = R_t+ R_{\Omega} \qquad $equation(\ref{15th_eqn})}
    \EndIf
    \State \text{observe $S_{i+1}$}
    \State \text{observe $R_t$}
    \State \text{update system with UAVs new  locations}
    \State \text{save $(S_i, A_i,r_i,S_{i+1})$ in replay memory}
    \State \text{sample a minibatch of $(S_i, A_i,r_i,S_{i+1})$}
    \State \text{$\theta_{old} \gets \theta$}
    \State \text{$error = R_t + \gamma V(S_{i+1}) - V(S_i)$}
    \EndIf
    \EndIf
\EndFor
\EndFor
\EndFor
\EndFor
\end{algorithmic}
\end{algorithm}

To stabilize our model's training using the PPO algorithm and ensure the convergence of the system model, we used a variety of methods and techniques summarized in Algorithm \ref{alg:cap}. In this algorithm, we initialize the CNN network parameters that will be used in the training process, where two copies of the Q-network with the same neural network configuration and weights (lines 2-3). Then, we prepare the swarm of UAVs for random distribution in the grid after ensuring that these UAVs respect our two main path planning constraints, which are described in $Cons2$ in equation (\ref{13th_eqn}) (lines 4-7). The training process generates a series of episodes that repeatedly occur (lines 9-41). The training process executes several sequential episodes (lines 10-29), then the buffer stores the experience tuples of the trained episodes (lines 29-41). The agent generates a different set of discrete actions. The actions vector has two parts: the first is the allocation distribution of inferences, and the second is related to UAV path planning (line 14). Furthermore, if the constraints of equation \ref{13th_eqn} are valid (line 15), the agent checks the type of layers it is processing and calculates penalties related to latency (lines 16-36). Besides, if the chosen participant respects the constraints in equation (\ref{13th_eqn}), a step reward is added to the agent's rewards (lines 19-22). Following the selection of a valid UAV, a penalty is added to the total reward to represent the latency of processing that layer. The agent then examines the path planning constraints. If a UAV in the swarm does not coexist with another UAV, a reward is given to the agent (lines 23-25). A reward is also given to each UAV that passes through hot cells at each time step in order to encourage the swarm of UAVs to traverse these hot cells (lines 26-28). Following that, the new state and total reward are computed to update the system of the new environment (lines 37-38). To improve decision accuracy, we select random samples from the replay buffer (line  33) into a minibatch at each step. The saved values are used to update the network parameters (lines 34-35). The aforementioned mechanism is known as experience replay, and it is used to stabilize learning.

\section{Simulation Results and Analysis}
\label{Perfomrance_evaluation}
In this section, the evaluation of our system model is presented, and a comparison with other baseline and recently proposed models is conducted. To achieve a fair comparison, we applied the same configuration of the environment in different comparative methods. The same configurations include applying the same number and type of UAVs, the same size of the cells, the same number and type of requests, and the same number of cells. Parameters are described in the following paragraph. Furthermore, we evaluate our system in terms of the accuracy of respecting different constraints, the latency of making final classification decisions, shared data, and the QoS of the system. In terms of QoS, we compare our results with two state-of-the-art works, which use the sigmoid demand algorithm and linear demand algorithm. The latency is defined as the delay from capturing an image from the ground until making the final classification decision.
The shared data is the total data transmitted between devices in the swarm. Moreover, the accuracy is defined as the percentage of respecting the constraints of the two tasks (path planning and layers allocations) after the convergence. Finally, the QoS of our system model is defined as the level of visiting the hot cells and providing services to the hot cells.

Our system covers a surveillance scenario. In the simulation testing, we used three different CNN networks, a small CNN network, namely LeNet (2 convolutional layers and 3 fully connected layers), which is trained with 32×32×3 RGB sized image, a medium-sized CNN network, namely AlexNet (5 convolutional layers and 3 fully connected layers), which is trained with a 227×227×3 RGB-sized image, and finally, the VGG16 CNN network (13 convolutional layers and 3 fully connected layers) which is trained with a 224×224×3 RGB-sized image. The UAVs are endowed with a camera to capture an image from the monitored area and launch the inference request.

We adopted an area of 100m $ \times$ 100m to be covered by the connected participants, in which we used 25 cells with 20m width and 20m length. Furthermore, we used three different types of UAVs that belong to the family of Raspberry Pi B+ \cite{daryanavard2018implementing}. Particularly, all types are characterized by a 1.4 GHz 64-bit quad-core processor and 1 GB RAM while they have a different number of multiplications per second $e_i$ \cite{disabato2019distributed} which are 560,512, and 256. We emphasize that our classification request generation follows a Poisson process. The algorithm of our system model is validated using the parameters summarized in Table \ref{calculation_data_summary}. We emphasize that these parameters are empirically chosen, and we believe that similar systems will perform identically.

\begin{table}[!t]
    \caption{Simulation Parameters.}
    \label{calculation_data_summary}
    \centering
    \begin{tabular}{|p{2cm}|p{3cm}|p{2cm}|}\hline
        Parameters & Description & Value \\\hline
       $\gamma$ & Discount factor &  0.99\\\hline
       $\alpha$ & Learning rate & 0.00025 \\ \hline
       $M$ & Batch size & 512 \\ \hline
       Policy & PPO2 algorithm & MlpPolicy \\ \hline
       $\epsilon$ & Exploration rate & 1  \\ \hline
       $\lambda$ & PPO coefficient & 0.95 \\ \hline
       $\varepsilon$ & Clip range & 0.2 \\ \hline
        $n_{epochs}$ & Epochs number & 4 \\ \hline
        $\sigma^2$ & Thermal noise power & $7.9 \times 10^{-9}$\\ \hline
        $P_{i,k}^t$ & UAV’s transmission power & 0.1 W\\ \hline
        $B_{i,k}^t$ & The allocated Bandwidth & 1 KHz\\ \hline
    \end{tabular}
\end{table}

\begin{figure}[!t]
    \centering
    \includegraphics[width=\linewidth]{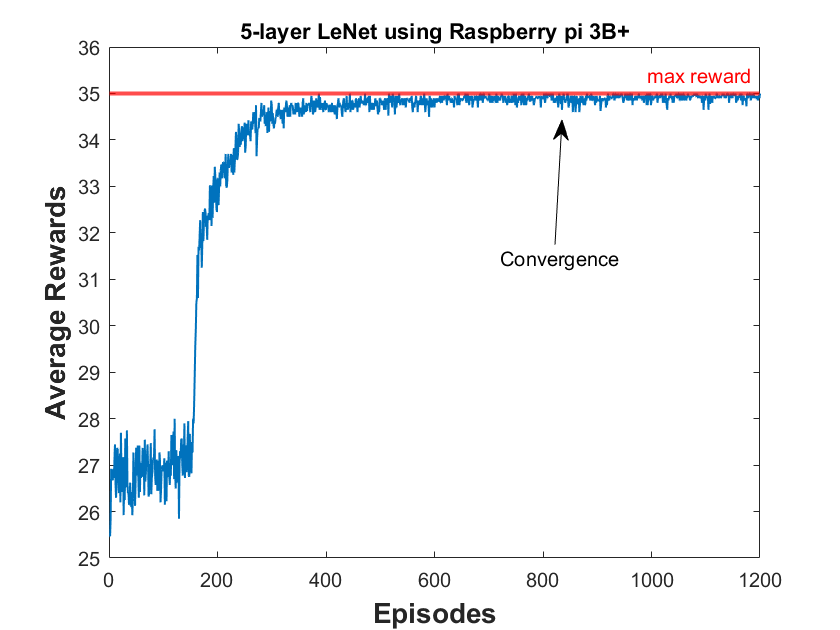}
    \caption{Average Cumulative Rewards for LeNet.}
    \label{fig:graph4}
\end{figure}

\begin{figure}[!t]
    \centering
    \includegraphics[width=\linewidth]{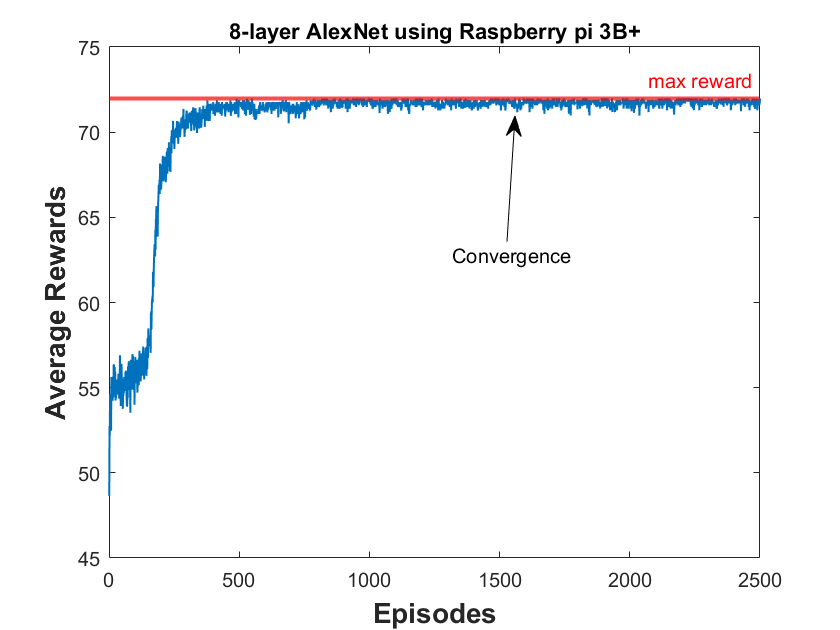}
    \caption{Average Cumulative Rewards for AlexNet.}
    \label{fig:graph5}
\end{figure}

\begin{figure}[!t]
    \centering
    \includegraphics[width=\linewidth]{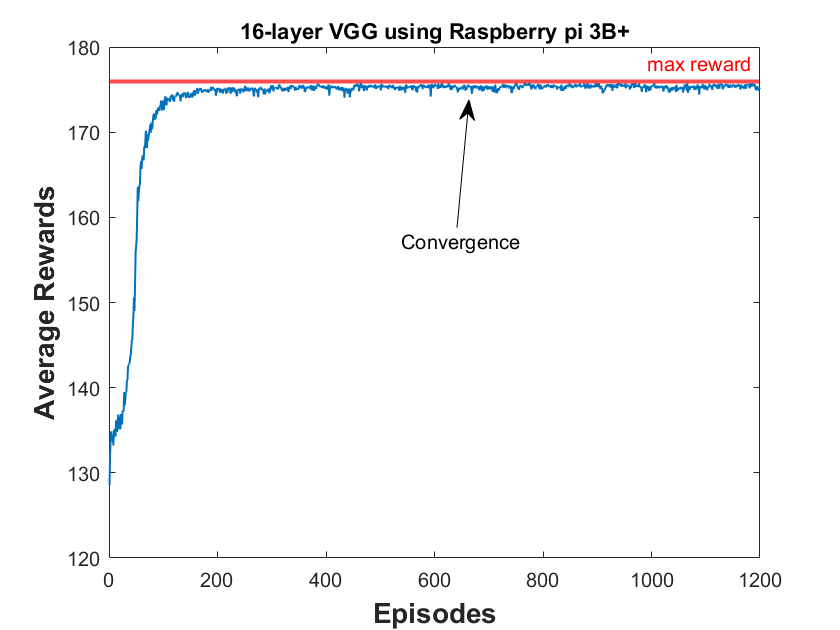}
    \caption{Average Cumulative Rewards for VGG16.}
    \label{fig:graph6}
\end{figure}

\begin{figure}[!t]
    \centering
    \includegraphics[width=\linewidth]{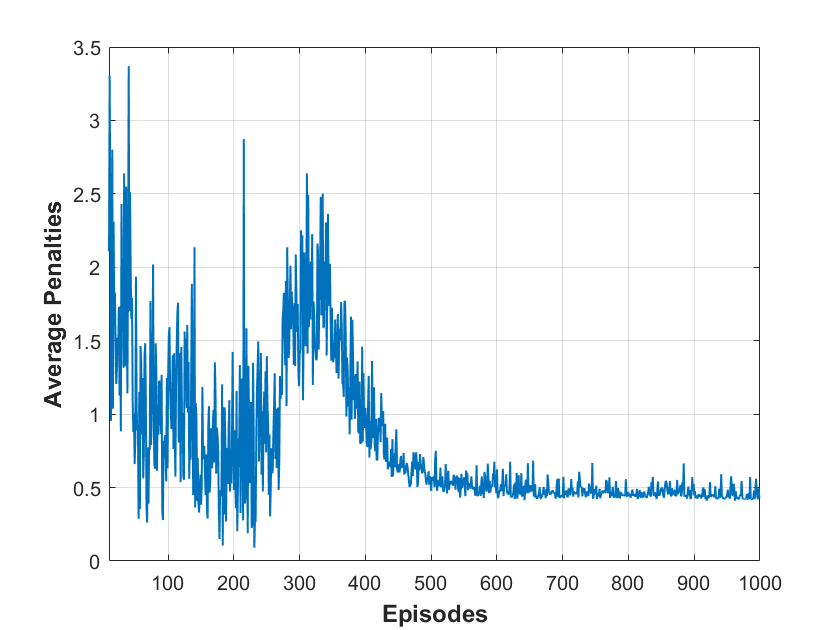}
    \caption{Average Penalties.}
    \label{fig:graph7}
\end{figure}

Figures \ref{fig:graph4}, \ref{fig:graph5}, and \ref{fig:graph6} show the result of the variation of the cumulative rewards over the training episodes of LeNet, AlexNet, and VGG16, respectively.
As shown in the figures, the first process is the exploration, where the agent starts learning the optimal policy and then starts increasing until it converges to the maximum cumulative reward. Moreover, VGG16 requires fewer episodes to converge and learn the optimal policy of executing the process and doing the multi-task operation because the episode length in VGG-16 is longer than the small CNN network such as LeNet, and the medium CNN network, AlexNet. We added to these graphs the line of maximum reward that the RL policy tries to pursue. This line is impossible to reach as we have a penalty related to latency. However, we can see that our algorithm can accomplish near-optimum results, and our system model is performing well. Figure \ref{fig:graph7} shows the convergence of penalties. In other words, it shows the main objective of the RL, which is minimizing the latency of calculating the classification decision. It starts to be high; then, the system learns how to decrease the penalty of calculating the latency until converging to the minimum that the swarm of UAVs can achieve.

\begin{figure}[!t]
    \centering
    \includegraphics[width=\linewidth]{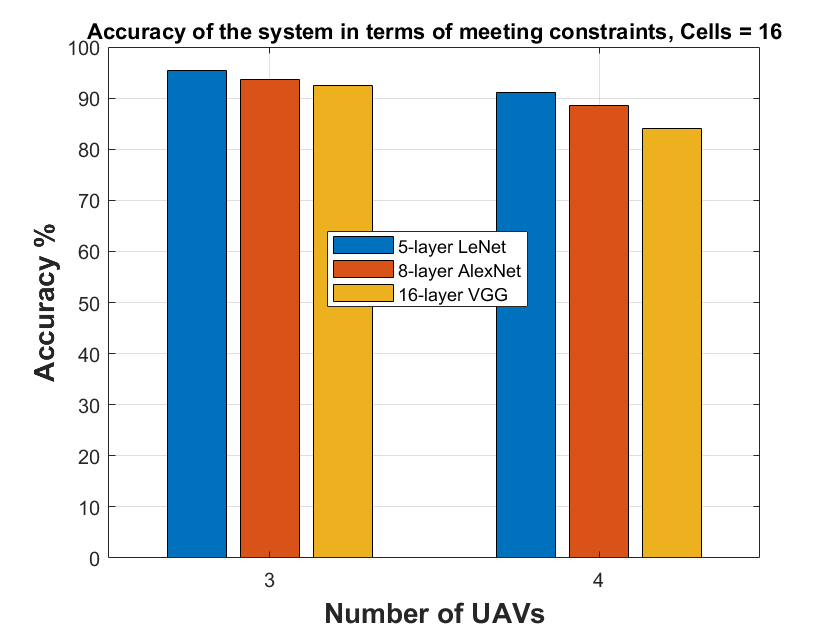}
    \caption{Accuracy comparison of different UAVs}
    \label{fig:graph8}
\end{figure}

In Figure \ref{fig:graph8}, we present the accuracy graphs for different CNN models. In these graphs, we show the capacity of respecting the constraints, calculated after the convergence, which means after the system learns the optimal allocation as well as the path planning of the connected UAVs that leads to the minimum latency of executing the final classification decisions. The simulated results were conducted in three different CNN networks, 5-layers LeNet, 8-layers Alexnet, and 16-layers VGG. The accuracy percentage of LeNet is 96.45\%, AlexNet is 91.23\%, and VGG16 is 82.50\%.

Due to the np-hardness of the problem in equation (\ref{8th_eqn}), it is difficult to calculate the optimal solution; hence, we evaluated the problem by calculating the sub-optimal solution of the main objective function. To calculate the sub-optimal solution, we fixed one variable and run the optimization problem for the other one, and we tried to run it for the maximum of possible values. Then, the values that give the minimum latency are chosen.

\begin{figure}[!t]
    \centering
    \includegraphics[width=\linewidth]{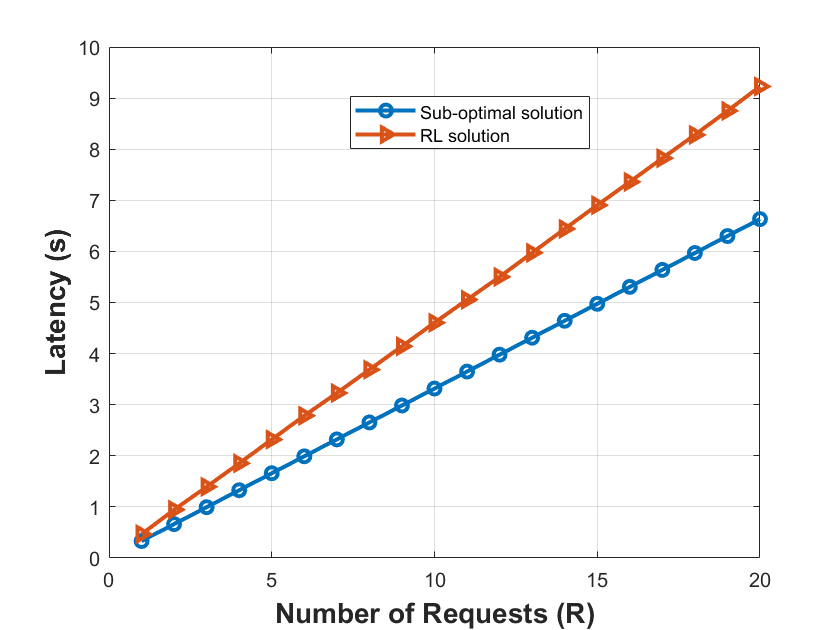}
    \caption{Comparing Optimal and RL solutions in terms of Latency.}
    \label{fig:graph9}
\end{figure}

\begin{figure}[!t]
    \centering
    \includegraphics[width=\linewidth]{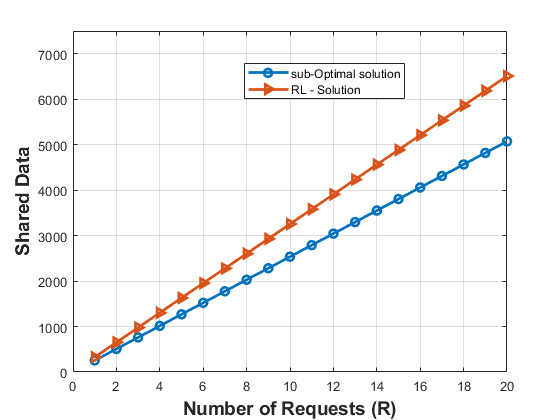}
    \caption{Comparing Optimal and RL solutions in terms of Shared Data.}
    \label{fig:graph10}
\end{figure}

Figures \ref{fig:graph9} and \ref{fig:graph10} compare the solution of the sub-optimal and RL in terms of cumulative latency and shared data. We adopt 5-layer LeNet with Raspberry pi 3B+ and 5 UAVs in this evaluation. From the figures, the difference between the two solutions is negligible at the beginning; it increases as the number of requests increases to a max of 40\% at 25 requests which is an extreme scenario that may not be realistic in the case of using 5 UAV devices.

\begin{figure}[!t]
    \centering
    \includegraphics[width=\linewidth]{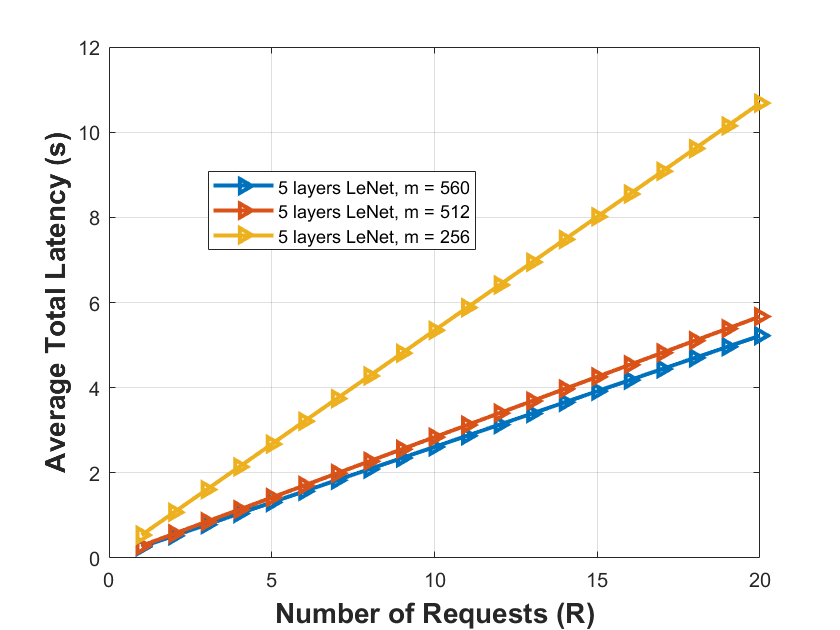}
    \caption{Average total latency of batches of requests for different types of devices.}
    \label{fig:graph11}
\end{figure}

Figure \ref{fig:graph11} shows the total latency of batches of requests when using Raspberry pi devices with different memory capacities. As shown in the figure, as the memory capacities of the devices increase, the latency of the inference decreases. For example, the latencies of the devices when $e$ = 512 and $e$ = 560 are quite similar when the number of requests is low, and the difference starts to be notified when the number of requests increases and the load becomes higher. The memory capacity becomes challenging at the higher number of requests, and the devices need to exploit all their resources to handle all requests. Furthermore, we can see that the average total latency of 20 requests for the lowest device capacity is 10.68 seconds. It means that the latency per request is equal to 0.53 seconds. When having 10 requests, the average latency is 5.34 seconds. For the highest device capacity, for 20 requests, the average total latency is 5.22 seconds, where the average latency per request is 0.26 seconds, and in the case of having 10 requests, the total average latency is 2.6 seconds. This proves the capacity of our system to run inference requests with low latency.
\begin{figure}[!t]
    \centering
    \includegraphics[width=\linewidth]{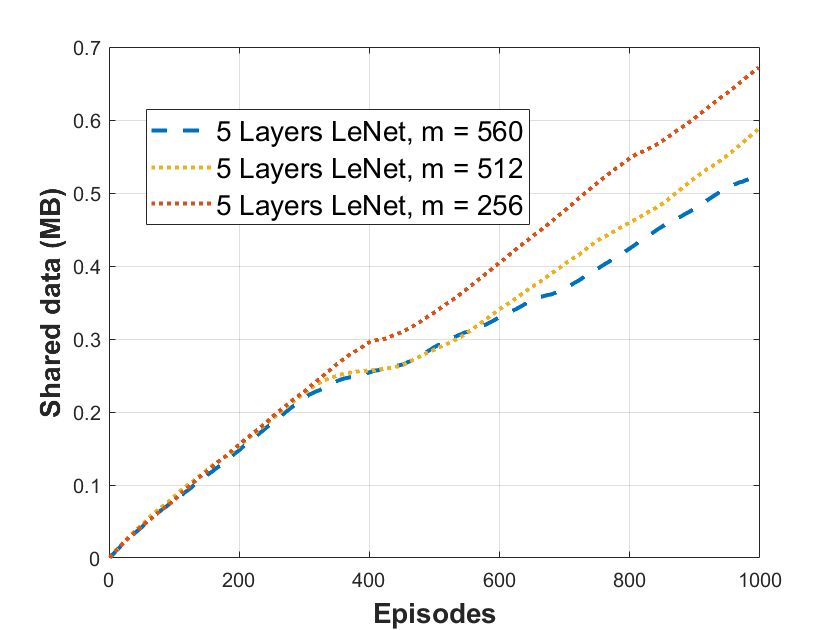}
    \caption{cumulative Shared Data among devices with different memory capacities.}
    \label{fig:graph12}
\end{figure}
\begin{figure}[!t]
    \centering
    \includegraphics[width=\linewidth]{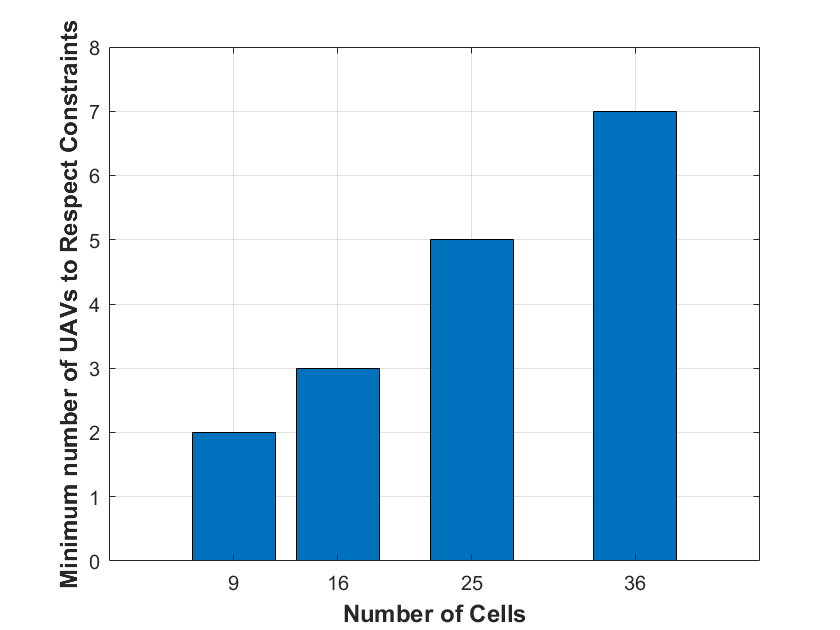}
    \caption{Minimum number of UAVs to respect constraints, LeNet distribution using Raspberry pi 3B+ device with 5 requests.}
    \label{fig:graph13}
\end{figure}
Figure \ref{fig:graph12} shows the shared data between UAV devices in the swarm. In this simulation, we compare different Raspberry Pi devices while varying the capacity of the memory, and we adopt AlexNet as the network used for classification. It is clear that as the memory capacity increases, the system incurs less shared data. This result is related to the incapability of the UAVs with low memory capacity to accomplish the whole classification on-board; hence they resort to distributing a part of the inference tasks.

Our approach contains two essential tasks (layers allocation and path planning). In order to satisfy these two tasks, the number of UAVs should be sufficient to complete their tasks successfully, and there is a minimum number of UAV devices in which lower than this number (threshold), the mission cannot be completed or completed with high latency and/or high interference due to their available resources. In Figure \ref{fig:graph13}, we present the minimum number of UAVs to start respecting the two constraints. Moreover, the minimum number of UAVs required to satisfy the mission depends on the number of cells. Clearly, the larger the number of cells, the more UAVs are required to achieve the mission. 
\begin{figure}[!t]
    \centering
    \includegraphics[width=\linewidth]{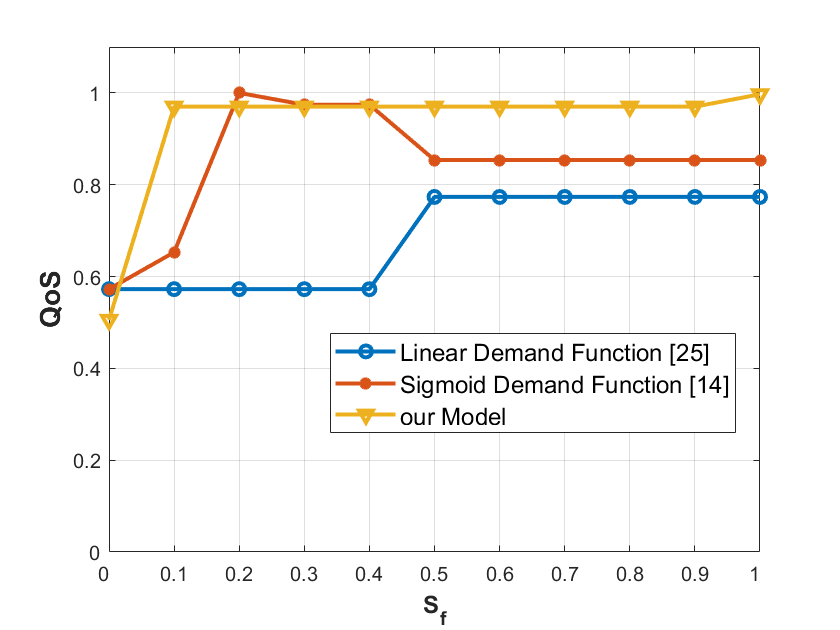}
    \caption{QoS simulation results.}
    \label{fig:graph14}
\end{figure}
\begin{figure}[!t]
    \centering
    \includegraphics[width=\linewidth]{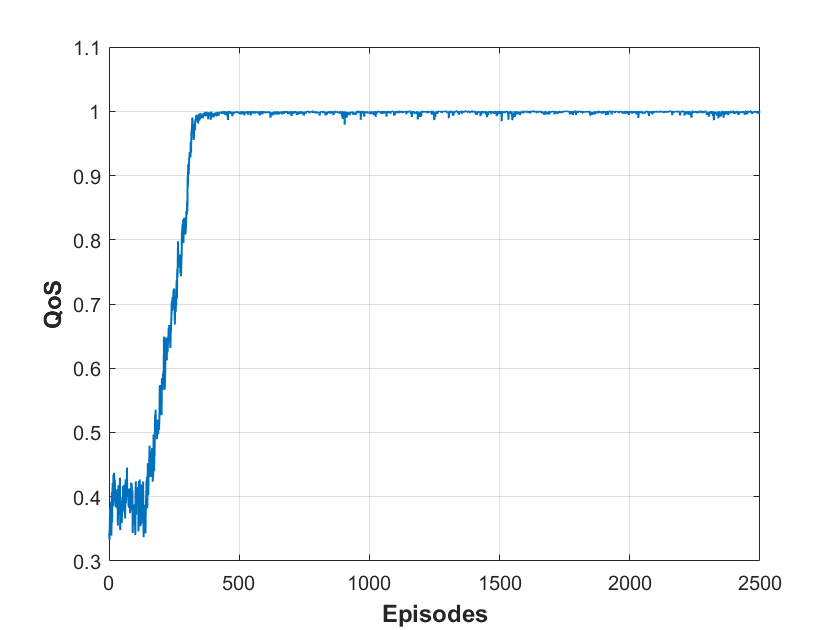}
    \caption{Convergence performance of rewards related to QoS for AlexNet, $S_f = 1$.}
    \label{fig:graph15}
\end{figure}

Figures \ref{fig:graph14} and \ref{fig:graph15} depict the QoS. The QoS is defined as how many times the UAVs visit the hot cells and provide services (capture requests and send them for classification) to these cells. The QoS coefficient factor $S_f$ encourages the agent to visit these hot cells by increasing the total reward that is based on equation (\ref{15th_eqn}). Figure \ref{fig:graph14} shows that when the QoS coefficient factor $S_f$ is 0, i.e., there is no reward for covering hot cells, our RL-agent gives a result of around(0.56); however, when $S_f=1$, the agent knows that there is a reward from traversing through these hot cells, so it encourages the UAVs to pass through these cells and provide services to them. We compared our result with two baseline works \cite{chang2021multi,liu2020path}, and we can see that when the QoS coefficient increases, our model does better than the other two models and provide services to all the hot cells in the ground. In these works, the authors used multi-UAVs to provide services to ground users, and these users have demand requests that need to be accomplished while UAVs plan their trajectory path to arrive at their destinations. The study in \cite{chang2021multi} used a sigmoid demand algorithm to fulfill the services of the users, while the study in \cite{liu2020path} used a linear demand algorithm. Our work outperforms these two works in which we assumed that we have hot cells, and inside these hot cells, demand services are needed to be fulfilled; hence, the UAVs traverse through these cells and provide services. Our work is based on the reward that is given to the agent, and when this reward is higher, the agent encourages the connected UAVs to visit these cells; therefore, when the services factor of equation (\ref{15th_eqn}) is high, the agent learns the optimal policy and cover these hot cells. Figure \ref{fig:graph15} shows the convergence of rewards related to QoS when using Raspberry pi with $e$ = 560. We can see that the model starts learning from past experiences until it converges to 1, which means that the swarm is able to provide services to all hot cells.

\begin{figure}[!t]
    \centering
    \includegraphics[width=\linewidth]{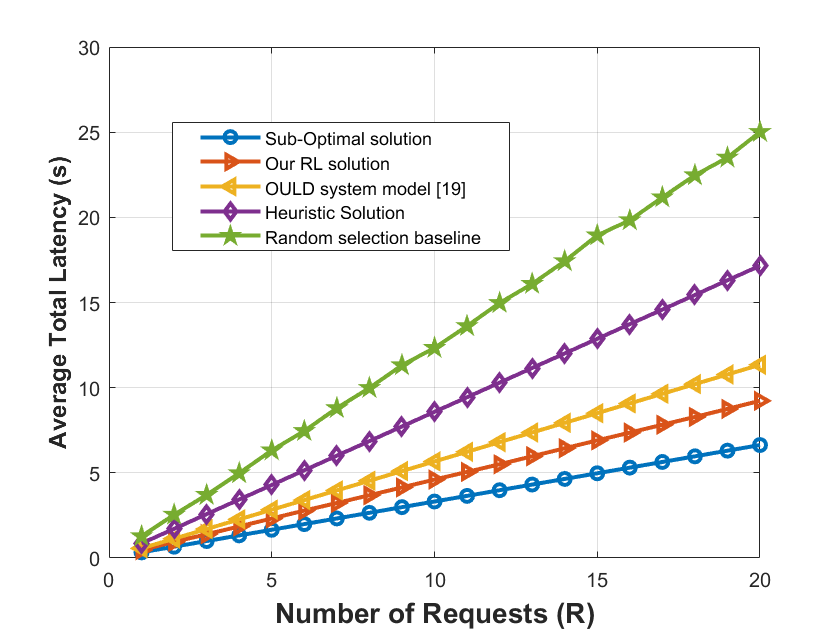}
    \caption{Total Latency of different models}
    \label{fig:graph16}
\end{figure}

Figure \ref{fig:graph16} shows a comparison of our system to the approach in \cite{jouhari2021distributed}, namely Optimal UAV-based Layer Distribution (OULD). In this model, the authors minimize the latency of the final prediction of the captured requests after assigning CNN layers; however, the paths of UAVs are static, and they are already planned for the UAVs at the initialization of the system. In our approach, the UAVs' paths are optimized to minimize the latency of the final prediction. For a fair comparison, we applied the same deep RL algorithm, the PPO algorithm, with the same input parameters to both system models. It is clear that our system is outperforming OULD in terms of latency as the paths of UAVs are optimized to find the trajectory that produces the least latency.
Moreover, we compare our approach to two different baselines, the greedy heuristic and the random selection strategy. In the heuristic solution, the paths are static, and the UAVs' availabilities are checked. The one that provides the least latency and has enough resources to execute the subsequent layer will be chosen. We note that the requests are served one by one greedily. The main difference between the heuristic solution and the OULD approach is that the RL algorithm in OULD  has the knowledge about the whole path that UAVs follow; however, the heuristic solution has the knowledge of only the next location where the UAV will fly; therefore the latency OULD system is better than the latency produced by the heuristic solution. In the random selection baseline approach, the UAVs randomly move, and the result of average latency is the worst due to the random movements of UAVs, and hence it is not adequate for online transmissions. Compared to the OULD system, the heuristic solution, the random selection, our system optimizes the UAV's movements to minimize the delay of the captured requests; hence it accomplishes the best results in terms of latency.

\begin{figure}[!t]
    \centering
    \includegraphics[width=\linewidth]{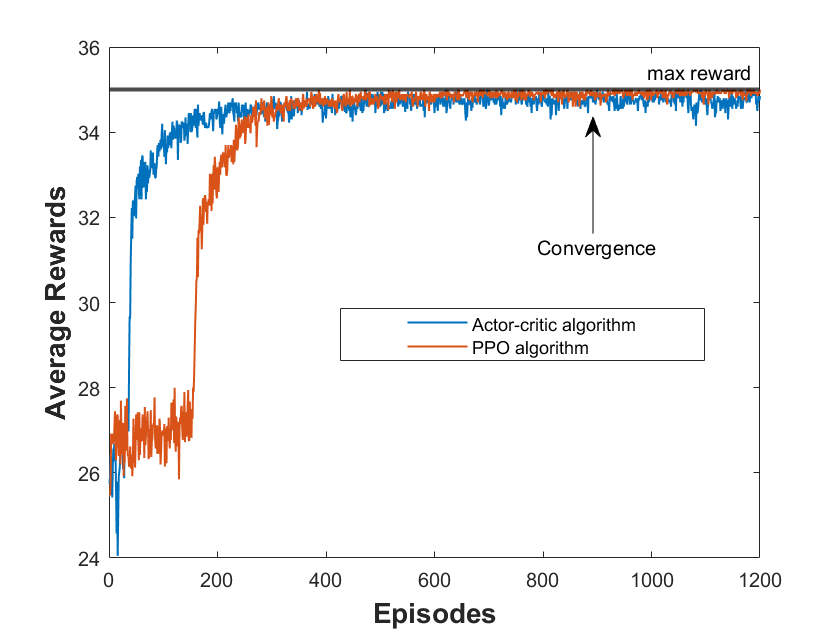}
    \caption{Accuracy comparison of different algorithms}
    \label{fig:graph17}
\end{figure}

In Figure \ref{fig:graph17}, we compare our PPO algorithm to a commonly used RL algorithm, namely the Actor-Critic algorithm. Both algorithms converge to the maximum reward represented by the black horizontal line in the figure. As we can see, the PPO algorithm outperforms the Actor-Critic algorithm in terms of optimality, even though the Actor-Critic converges faster. Clearly, the PPO algorithm takes more time to learn the optimal layers allocation and plan the path of UAVs; however, it learns a better policy compared to the Actor-Critic. Finally, the accuracy of respecting the constraints of PPO is 94\%, whereas Actor-Critic achieves 88\%.
\begin{figure}[!t]
    \centering
    \includegraphics[width=\linewidth]{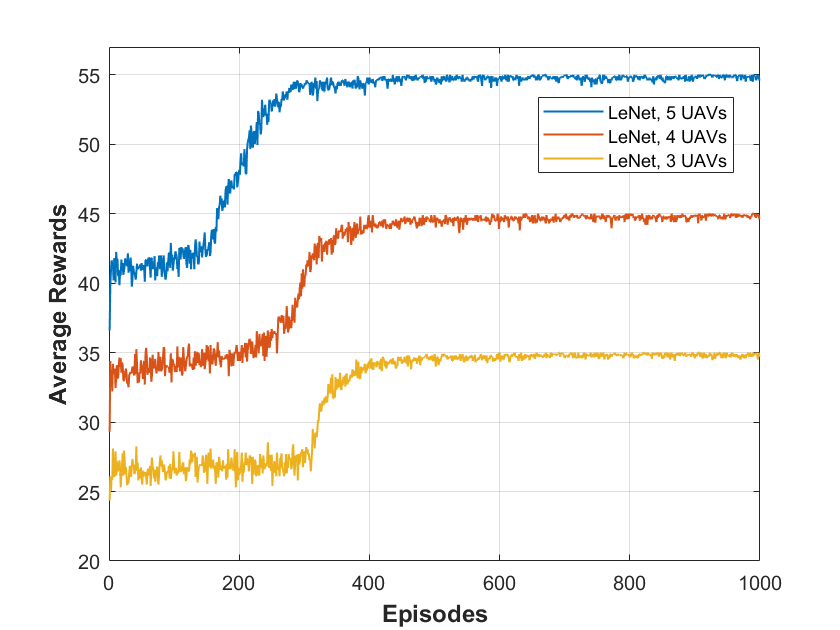}
    \caption{Average cumulative rewards when varying the number of UAVs.}
    \label{fig:graph18}
\end{figure}
\begin{figure}[!t]
    \centering
    \includegraphics[width=\linewidth]{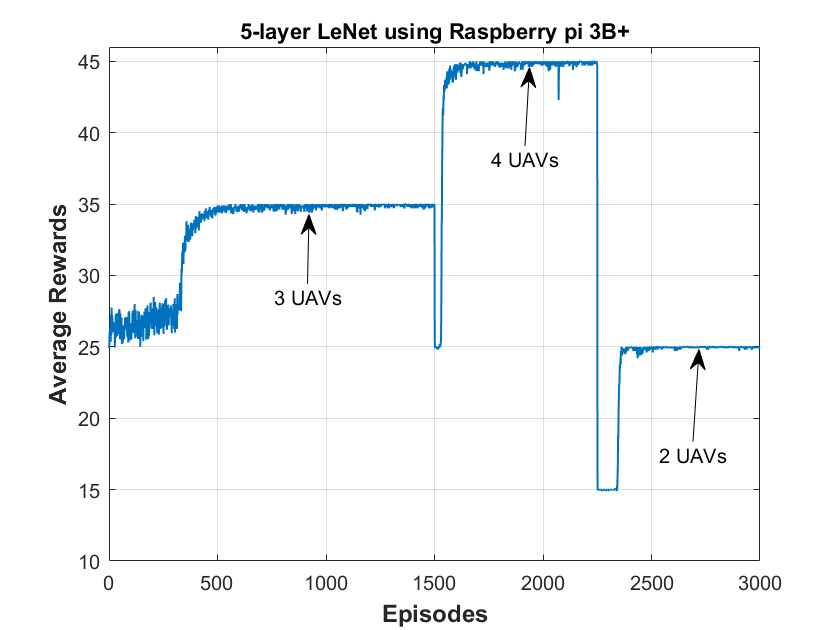}
    \caption{Average cumulative rewards under a dynamic environment.}
    \label{fig:graph19}
\end{figure}

Figure \ref{fig:graph18} refers to the comparison between swarms enclosing different numbers of UAVs in terms of average cumulative rewards. As the number of UAVs increases, the convergence to the maximum rewards becomes faster as the agent has enough choices to choose from among the available UAVs. In Figure \ref{fig:graph19}, we test changing the environment of the learning. This can help us test the prompt adaptability of our system to any changes knowing that the number of UAVs can be increased or decreased at any time to go to the charging station, for example, and rejoin the swarm after a period of time when the battery is fully charged. Particularly, the system starts with three UAVs, and it converges to the maximum rewards. After converging, one UAV is assumed to join the swarm. The accuracy is decreased as the agent tries to learn the new environment, then it can re-converge rapidly to the maximum rewards where 4 UAVs are deployed. Decreasing the number of UAVs shows the same performance. More specifically, the cumulative rewards drop when the configuration of the system changes and then increase again after a small learning period.
\section{conclusion}
\label{Conclusion}
In this paper, we introduced a distributed system model of a swarm of UAVs while planning the trajectory path of UAVs. The system model is designed to overcome the limitations of classifying the captured image in remote servers by disrupting the task into segments and executing these segments in the connected UAVs by exploiting the limited capacities of UAVs. Simultaneously, the UAVs draw their path to cover the monitored region and focus on hot cell locations to capture requests. We formulated the distributed system model as an optimization problem that seeks to minimize the latency of making the final classification decision while finding the best path that the UAV can pass through to mitigate interference. The formulated problem is np-hard, and it is not adequate for real-time applications. Therefore, we introduced an online solution based on the RL algorithm that is possible for dynamic models and suitable for real-time applications. Our simulation results demonstrated the effectiveness of our proposed solution to minimize the latency and improve the accuracy of respecting the constraints. In our future work, we plan to study the security of transmitting data as these applications require high-security transmission and achieve high reliability. Moreover, we plan to study the effect of interference on transmission latency when controlling channel communication under different modes like a line of sight and a non-line of sight. Furthermore, we plan to investigate the Doppler effect of moving UAVs as it affects the interference of transmission between moving objects. 
\bibliographystyle{IEEEtran}
\bibliography{bibliography.bib}
\begin{IEEEbiography}[{\includegraphics[width=1in,height=1.25in,clip,keepaspectratio]{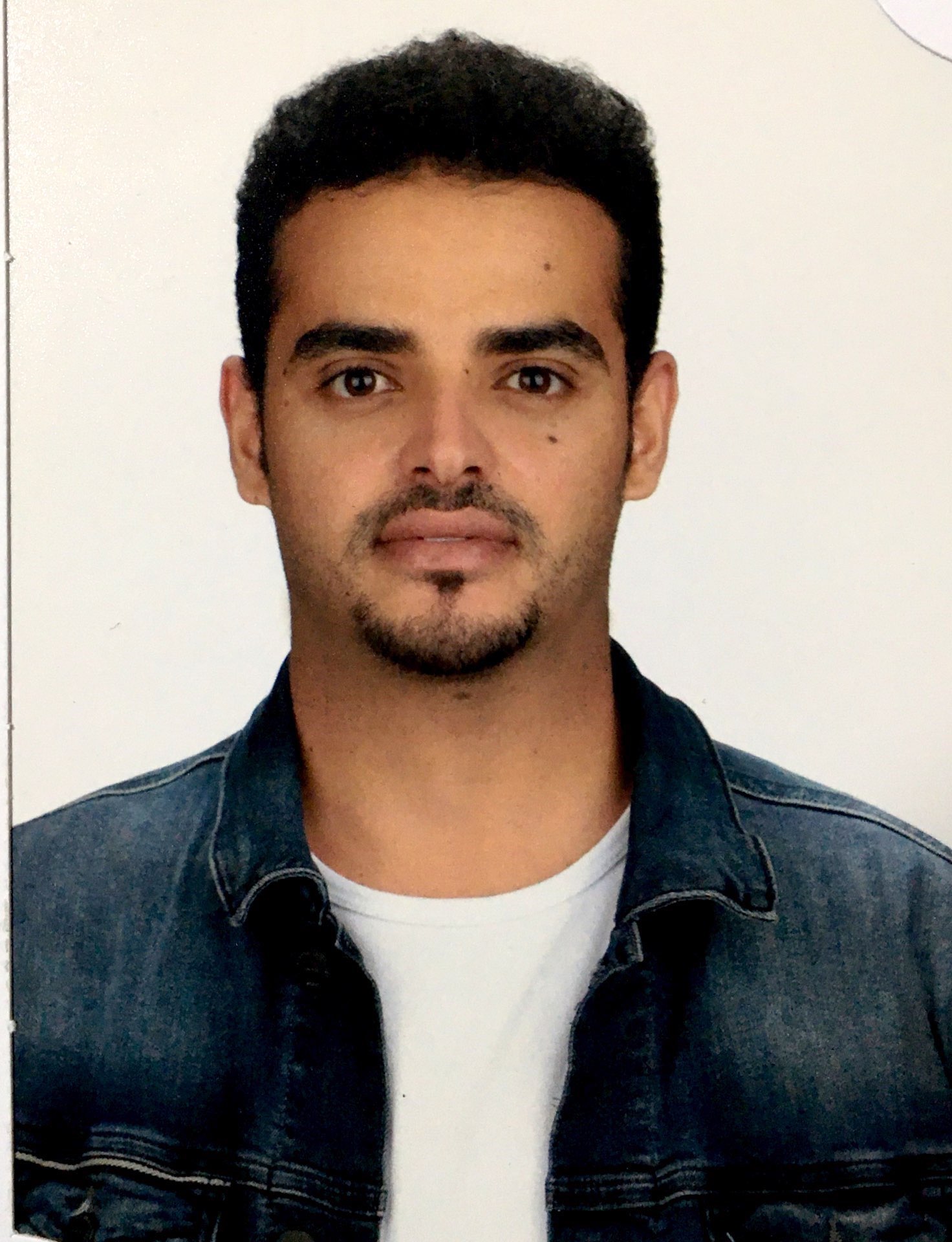}}]%
{Marwan Dhuheir}
received the master’s degree from Kocaeli University - Turkey, in 2019. He is currently pursuing the Ph.D. degree with the Department of Information and Computing Technology (ICT), Hamad Bin Khalifa University (HBKU), Doha – Qatar. His research interests include machine learning, unmanned aerial vehicles, wireless communications, deep reinforcement learning, and edge computing.
\end{IEEEbiography}
\begin{IEEEbiography}[{\includegraphics[width=1in,height=1.25in,clip,keepaspectratio]{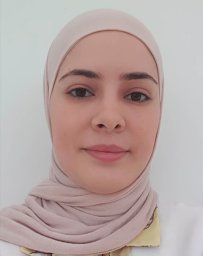}}]%
{Emna Baccour}
received the bachelor’s degree in computer science networks and telecommunications from the National Institute of Applied Science and Technology, in 2011, the master’s degree in electronic systems and communication networks from the Polytechnic School, in 2012, and the Ph.D. degree in computer science from the University of Burgundy Dijon, France, in 2017. She was a Research Assistant at Qatar University on a project covering the interconnection networks for massive data centers, where she currently holds a postdoctoral position. Her research interests include data center networks, edge computing, green computing, machine learning, and deep reinforcement learning techniques.
\end{IEEEbiography}
\begin{IEEEbiography}[{\includegraphics[width=1in,height=1.25in,clip,keepaspectratio]{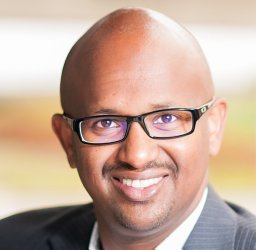}}]%
{Aiman Erbad (Senior Member, IEEE)}
received the M.Sc. degree from the University of Essex, in 2005, and the Ph.D. degree from The University of British Columbia, in 2012. He is currently an Associate Professor with the College of Science and Engineering, Hamad Bin Khalifa University (HBKU). His research interests include cloud computing, edge computing, the IoT, private and secure networks, and multimedia systems. He received the Platinum Award from H. H. Emir Sheikh Tamim bin Hamad Al Thani at the Education Excellence Day 2013 (Ph.D. category). He also received the 2020 Best Research Paper Award from Computer Communications, the IWCMC 2019 Best Paper Award, and the IEEE CCWC 2017 Best Paper Award. His research interests include cloud computing, edge computing, the IoT, distributed AI algorithms, and private/secure networks. He is currently an Editor of the International Journal of Sensor Networks (IJSNet) and KSII Transactions on Internet and Information Systems, and served as a Guest Editor for IEEE Networks.
\end{IEEEbiography}
\begin{IEEEbiography}[{\includegraphics[width=1in,height=1.25in,clip,keepaspectratio]{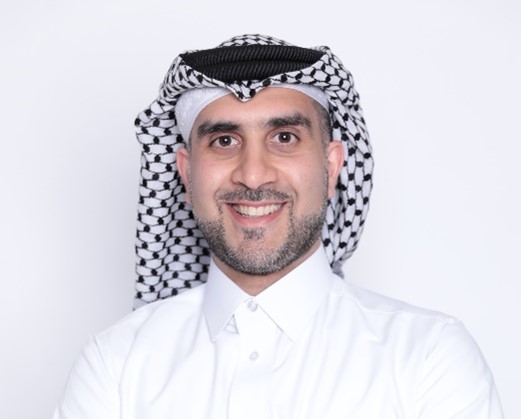}}]%
{Sinan Sabeeh Al-Obaidi}
received his PhD and MEng from Electrical Engineering Department at Texas A\&M University at College Station, TX through a scholarship offered by Qatar Research and Leadership Program from Qatar Foundation and his BSc in Electrical Engineering from Texas A\&M University at Qatar through an honorary scholarship from Her Highness Sheikha Moza Bint Nasser Al-Missnad. Currently Dr. Al-Obaidi directs a portfolio of projects in autonomous systems as the Manager of Engineering in R\&D directorate at Barzan Holdings QSTP LLC for defense and security applications.
\end{IEEEbiography}
\begin{IEEEbiography}[{\includegraphics[width=1in,height=1.25in,clip,keepaspectratio]{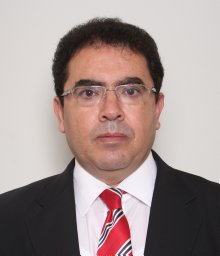}}]%
{Mounir Hamdi, (Fellow, IEEE)}
is currently the Founding Dean of the College of Science and Engineering, Hamad Bin Khalifa University (HBKU). He is an IEEE Fellow for contributions to design and analysis of high-speed packet switching. As the Founding Dean of the College of Science and Engineering, he led the foundation of 15 graduate programs and one bachelor’s program and all the associated research laboratories and activities. Before joining HBKU, he was the Chair Professor at The Hong Kong University of Science and Technology (HKUST), and the Head of the Department of Computer Science and Engineering. From 1999 to 2000, he was a Visiting Professor at Stanford University, USA, and the Swiss Federal Institute of Technology, Lausanne, Switzerland. His research interest includes high-speed wired/wireless networking in which he has published more than 400 research publications, received numerous research grants, and graduated more 50 M.S./Ph.D. students. He is/was on the Editorial Board of various prestigious journals and magazines, including IEEE TRANSACTIONS ON COMMUNICATIONS, IEEE Communication Magazine, Computer Networks, Wireless Communications and Mobile Computing, and Parallel Computing. In addition to his commitment to research and professional service, he is also frequently involved in higher education quality assurance activities and engineering programs accreditation all over the world.
\end{IEEEbiography}
\end{document}